\documentclass[%
superscriptaddress,
amsmath,amssymb,
aps,
prd,
]{revtex4-2}

\usepackage[colorlinks,linkcolor=blue,citecolor=magenta,urlcolor=blue]{hyperref}
\usepackage{bm,graphicx,dcolumn,epstopdf,epsf, latexsym,mathbbol, amssymb,amsmath,color,slashed, mathrsfs,mathcomp,simplewick}
\usepackage{graphicx}
\usepackage{tabularx}
\usepackage{dcolumn}
\usepackage{bm}
\usepackage{orcidlink} 
\usepackage{graphicx}
\usepackage{hyperref}

\usepackage{booktabs}  

\newcommand{\orcid}[1]{%
  \href{https://orcid.org/#1}{%
    \includegraphics[width=10pt]{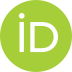}%
  }%
}
\usepackage{array}
\usepackage{booktabs}
\usepackage{tabu}
\usepackage{dcolumn}

\usepackage{amsmath}
\usepackage{amsfonts}
\usepackage{amssymb}
\usepackage{graphicx}
\usepackage{subfigure}
\usepackage{graphicx}
\usepackage{bm}
\usepackage{xcolor}
\usepackage{adjustbox}  

\graphicspath{{figs/}} 

\begin{document}

\title{A Starobinsky-like Potential in k-Essence Framework: Attractor Dynamics, Reheating, and Consistency with ACT DR6}

\author{Abolhassan Mohammadi\orcid{0000-0003-1228-9107}}
    \email{abolhassanm@hnit.edu.cn}
    \affiliation{%
School of Physics, Hunan Institute of Technology, Hengyang, China.
}%
\author{Yogesh\orcid{0000-0002-7638-3082}}%
    \email{yogeshjjmi@gmail.com}
    \affiliation{%
Centre for Cosmology and Science Popularization (CCSP), SGT University, Gurugram, Haryana 122505, India.
}%

\author{Hongwei Tan}
    \email{honweitan@hnit.edu.cn}
    \affiliation{%
School of Physics, Hunan Institute of Technology, Hengyang, China.
}%


\author{M. Sami\orcid{0000-0003-1081-0632}}%
    \email{sami$_$ccsp@sgtuniversity.org, samijamia@gmail.com}
    \affiliation{%
Centre for Cosmology and Science Popularization (CCSP), SGT University, Gurugram, Haryana 122505, India.
}%
    \affiliation{Eurasian International Centre for Theoretical Physics, Astana, Kazakhstan}
\date{\today}

\begin{abstract}
The recent ACT DR6 has shifted the preferred value of the scalar spectral index upward so that many well-established inflationary models have been disfavoured, including the Starobinsky potential. Despite this, the potential remains exceptionally well-motivated, with origins in $R^2$ gravity, no-scale supergravity, and the $\alpha$-attractor framework. In this work, we study a Starobinsky-like potential within a k-essence framework, described by the Lagrangian $\mathcal{L} = F(\phi)X - V(\phi)$, with a power-law kinetic coupling $F(\phi) = 1+A\phi^n$ and no modification to the gravitational sector, and show that this potential becomes consistent with ACT DR6 data. Solving the background numerically, we find that the predictions for $n_s$, $\alpha_s$, and $r$ fall within the $1\sigma$ region of ACT DR6 for a well-defined range of the coupling parameters. The attractor behavior of the inflationary solution is confirmed both analytically through the Hamilton-Jacobi formalism and numerically via a phase-space analysis. For the reheating phase, the nature of the potential fixes the effective equation of state to $w_{\rm re} = 0$, resulting in a reheating temperature $T_{\rm re} \sim 10^{14}~{\rm GeV}$, well above the BBN bound. We also compute the relic gravitational wave spectrum and find it within the BBO sensitivity bound. These results demonstrate that a Starobinsky-like potential is a viable candidate for inflation, and predictions of this type of plateau potential can be in agreement with ACT DR6 through a simple non-canonical kinetic coupling with no modification to the gravitational sector.

\end{abstract}

\maketitle


\section{Introduction}
\label{sec:intro}

In the early 1980s, the idea of inflation was introduced to solve the horizon and flatness problems of the early universe~\cite{starobinsky:1980te,Guth:1980zm,Albrecht:1982wi,Linde:1981mu,Linde:1983gd}. A brief period of exponential expansion that the universe went through during the first moments after the Big Bang. This naturally leads to a nearly flat, scale-invariant power spectrum.
Quantum fluctuations generated during the inflationary phase later become classical after the horizon entry and result in the large-scale structures ~\cite{Starobinsky:1982ee, Guth:1982ec, Baumann:2009ds}.  The two most important observables that must match the latest observations for any model to be compatible with the data are the tensor-to-scalar ratio and the scalar spectral index. The viability of any inflationary model ultimately depends on its consistency with the latest cosmological observations~\cite{WMAP:2010qai, Planck:2018jri, Vagnozzi:2022qmc}. Observations from the WMAP and Planck satellites have imposed severe constraints on the inflationary model. Several models, including chaotic inflation, were ruled out by Planck. With recent developments in observational cosmology, the new observation from the Atacama Cosmology Telescope, ACT DR6~\cite{ACT:2025fju, ACT:2025tim}, has put several of the most widely accepted models of inflation in a tight spot. Notably, Starobinsky inflation arises from $R + R^2$ term in the effective action~\cite{starobinsky:1980te}. Furthermore, models like $\alpha-$attractors are also in tension with the ACT observations~\cite{Ellis:2025zrf}. A combined measurement of Planck and ACT leads to the higher value of $n_s = 0.9709 \pm 0.0038$, while incorporating baryon acoustic oscillation data from DESI (P-ACT-LB) pushes this further to $n_s = 0.9743 \pm 0.0034$ with an upper bound $r < 0.038$ at 95\% confidence (P-ACT-LB-BK18). This new bound on $n_s$ has forced us to look beyond the standard cold inflationary scenarios, and several attempts have already been made to extend or modify the models that are now disfavoured by the data~\cite{Kallosh:2025rni, Aoki:2025wld, Dioguardi:2025vci, Salvio:2025izr, Brahma:2025dio, Gao:2025onc, Drees:2025ngb, Zharov:2025zjg, Yin:2025rrs, Liu:2025qca, Gialamas:2025ofz, Haque:2025uri, Haque:2025uis, Dioguardi:2025mpp, Gialamas:2025kef, Qiu:2025iqm, Huang:2013hsb, Shobcha:2026mpc, Zahoor:2025nuq, Kim:2025dyi, Odintsov:2025eiv, Odintsov:2025wai, Sabogal:2026qvy, Bezerra-Sobrinho:2025gfg, Odintsov:2026dss, Oikonomou:2026mvp, Odintsov:2026doe,Oikonomou:2026vrw,Oikonomou:2025htz}. Only a handful of models exist that can pass the litmus test of the ACT observations while remaining within the premises of the standard cold inflationary scenario~\cite{Mohammadi:2025gbu}. Another way of keeping the model viable is embedding it in the modified gravity framework. It has been shown that modified gravity, such as the EGB, can save these models~\cite{Yogesh:2025wak}

In this work, we adopt a different route. We modify the kinetic sector of the scalar field while keeping Einstein's gravity this leads to the framework of $k-$inflation introduced by Armendariz-Pic\'on, Damour, and Mukhanov~\cite{Armendariz-Picon:1999hyi} and Garriga and Mukhanov~\cite{Garriga:1999vw}, where the scalar Lagrangian is a general function of the field $\phi$ and its kinetic term $X \equiv -\frac{1}{2}g^{\mu\nu}\partial_\mu\phi\,\partial_\nu\phi$. The broader theoretical setting of $k$-essence was laid out in~\cite{Armendariz-Picon:2000ulo}, and such Lagrangians arise naturally from low-energy limits of string theory and as sub-sectors of the general Horndeski scalar-tensor theory~\cite{Horndeski:1974wa}.

The specific model we consider is defined by the Lagrangian, $\mathcal{L} = F(\phi)\,X + V(\phi),$
where $F(\phi) > 0$ is a field-dependent coupling function. This form emerges from the Horndeski framework when the cubic Galileon term $G(\phi)\,\square\phi$ is integrated by parts, giving $F(\phi) = 1 - 2\,dG/d\phi$ \cite{Lin:2020goi, Solbi:2021rse}. Setting $F = 1$ recovers canonical inflation, so the model is a controlled deformation from a well-understood limit. Since the Lagrangian is linear in $X$, the sound speed of curvature perturbations equals unity~\cite{Garriga:1999vw}, so non-Gaussianity constraints are automatically satisfied. 

The coupling function $F(\phi)$ introduces an additional friction term in the inflaton equation of motion, modifying the slow-roll parameters and altering inflationary observables $(n_s, r)$ relative to the canonical case. For the coupling, we take a power-law form $F(\phi) \propto \phi^n$. The power index $n$ controls how strongly the non-canonical kinetic term modifies the slow-roll dynamics, giving the model enough freedom to fit the observations without unnecessary complexity. Similar models of $k-$inflation have been considered in the past in the context of primordial black holes~\cite{Lin:2020goi, Solbi:2021rse} 

For the potential, we take a potential that has a Starobinsky-like form 
\begin{equation}\label{eq:starobinsky}
    V(\phi) = V_0\left(1 - e^{-\sqrt{2/3}\,\phi}\right)^2,
\end{equation}
The Starobinsky potential arises from the $R + R^2$ modified gravity action, introduced in~\cite{starobinsky:1980te}, originally as a mean to solve the big bang singularity. After recasting the model in the Einstein frame, one obtains a canonical scalar field with a plateau potential. The potential also emerges in the context of the no-scale supergravity~\cite{Ellis:2025zrf}, where it can be realized as the special case $\alpha = 1$ of the broader $\alpha$-attractor framework~\cite{Kallosh:2013yoa}. Taken together, the Starobinsky can be addressed as one of the most well-motivated potential in inflationary cosmology. Despite these strong theoretical motivation for the potential, the ACT data release places it in tension with observations within the canonical single-field framework~\cite{ACT:2025fju, ACT:2025tim, Ellis:2025zrf}.

In addition to the observational predictions, such as the tensor-to-scalar ratio and the scalar spectral index, any inflationary model must be also not sensitive to the choice of the initial condition, in other word, and inflationary model should have attractor behavior. In such case, the inflationary trajectories in the phase-space of the scalar field, which are started from different initial field velocity, converge to the same slow-roll solution. Such behavior known as attractor behavior. Here, we consider the attractor behavior both analytically and numerically. For the analytical approach we use the Hamilton-Jacobi formalism \cite{Salopek:1992qy,Liddle:1994dx,Kinney:1997ne,Guo:2003nz}, and show that the small perturbation of the Hubble parameter will decay exponentially so that the coupling function control the rate of this decay. For the numerical case, we explore the phase-space trajectory of scalar field and its velocity \cite{Guo:2003nz,Remmen:2013eja}, and we show that the trajectories converge to the slow-roll solution, and have a spiral motion around the origin. We demonstrate that the attractor behavior is preserved in the presence of the field-dependent kinetic coupling $F(\phi)$.

The paper is organized as follows. Section~\ref{sec:kinfaltion} introduces the model and derives the field equations. Section~\ref{sec:results} presents the inflationary predictions and their comparison with ACT data, and also we discuss the attractor behavior of the solution. The reheating phase is discuss in Sec.\ref{sec:reheating}, where the reheating temperature and e-fold is estimated. Then, we study the primordial gravitational waves produced by the model in Sec.\ref{sec:gw}.  Section~\ref{sec:conclusion} summarizes our main findings. 

\section{k-inflation}
\label{sec:kinfaltion}

The k-essence model under consideration here can be regarded as a subclass of the G-inflation model. The G-inflation framework provides a theoretically complete and ghost-free setting for the non-canonical scalar field models~\cite{Deffayet:2010qz,Kobayashi:2010cm, Kamada:2010qe,Mohammadi:2023kzd,Mohammadi:2019qeu,Mohammadi:2015jka,Babichev:2007dw}. The action of G-inflation in general is given as
\begin{equation}\label{eq:G_action}
    S = \int d^4x \sqrt{-g} \left[ \frac{M_p^2}{2} R
    + K(\phi, X) - G(\phi, X)\,\Box\phi \right],
\end{equation}
where $M_p = (8\pi G_N)^{-1/2}$ is the reduced Planck mass, $g$ is the determinant of the metric $g_{\mu\nu}$, $R$ is the Ricci scalar, and $X \equiv -\frac{1}{2}g^{\mu\nu}\partial_\mu\phi\,\partial_\nu\phi$ is the kinetic term of the scalar field $\phi$. Here $K(\phi, X)$ and $G(\phi, X)$ are arbitrary functions of both the scalar field $\phi$ and the kinetic term $X$, and $\Box\phi \equiv \frac{1}{\sqrt{-g}}\partial_\mu\left(\sqrt{-g}\, g^{\mu\nu}\partial_\nu\phi\right)$ is the covariant d'Alembertian operator. It should be noted that the action~\eqref{eq:G_action} is itself a sub-category of the Horndeski scalar-tensor theory~\cite{Horndeski:1974wa}, which provides the most general scalar-tensor action that yields second-order field equation and therefore is free from the Ostrogradsky ghost instabilities. G-inflation is known as a theoretically well-motivated framework, in which simpler models, including ours, emerges from it as special cases. 

The model we are going to study here yields by applying two simplification to the action~\eqref{eq:G_action}. As the first simplification, the kinetic function is taken in its canonical form, i.e., $K(\phi, X) = X - V(\phi)$. Second, the coupling function $G$ is assumed to depend only on the scalar field $G = G(\phi)$. Under these simplification, it can be shown that the term $G(\phi)\,\Box\phi$ can be converted into the $-2\,G_{,\phi}(\phi)\,X$ through the integration by part, where $G_{,\phi} = dG/d\phi$ \cite{Lin:2020goi, Solbi:2021rse}. Therefore, by defining the function $F(\phi) \equiv 1 - 2G_{,\phi}(\phi)$, the action reduced to the k-essence action as
\begin{equation}\label{eq:action}
    S = \int d^4x \sqrt{-g} \left[ \frac{M_p^2}{2} R
    + F(\phi)\,X - V(\phi) \right].
\end{equation}
To avoid the ghost instability, the condition $F(\phi) > 0$ must be satisfied. In the limit $F = 1$, the action reduces to that of canonical single-field inflation. Then, the non-canonical coupling represents a well-defined and controlled extension of the standard case. Action~\eqref{eq:action} is a special case of k-essence, where the field-dependent coefficient $F(\phi)$ is multiplying the kinetic term $X$  while the gravity sector remains unchanged. A further theoretical motivation comes from the fact that Lagrangians of this type arise in low-energy effective actions of string theory~\cite{Armendariz-Picon:1999hyi, Garriga:1999vw}.

This class of models has been studied in different contexts. Their inflationary observables and agreement with CMB data were examined in~\cite{Barenboim:2007ii, Franche:2010yj}, while the perturbation theory and sound speed of this class of models were first worked out in~\cite{Armendariz-Picon:1999hyi, Garriga:1999vw}. Recently, the same Lagrangian has been used to study the formation of primordial black holes, in which an enhanced power spectrum was obtained by choosing a coupling with carefully adjusted parameters~\cite{Lin:2020goi, Solbi:2021rse}. Another notable feature of action~\eqref{eq:action} is that, because it is linear in $X$, the sound speed of curvature perturbations is exactly unity, $c_s = 1$~\cite{Garriga:1999vw}. 
Since the action is linear in the kinetic term $X$, one can remove the coupling by a field redefinition $d\psi/d\phi = \sqrt{F(\phi)}$ and bring the model back to the standard canonical scalar field model with a a different field variable and potential $V(\psi)$. For the rest of the work, we are going to keep working with the scalar field $\phi$ and coupling $F(\phi)$ because this is the frame where the coupling function has a clear origin from the Galileon term as explained above. 

\subsection{Background equations}
\label{subsec:background}

We assume that the geometry of the universe is described by a spatially flat FLRW metric,
\begin{equation}\label{eq:metric}
    ds^2 = -dt^2 + a^2(t)\,\delta_{ij}\,dx^i\,dx^j,
\end{equation}
where $a(t)$ is the scale factor. Varying action~\eqref{eq:action} with respect to the metric $g_{\mu\nu}$ and applying metric~\eqref{eq:metric}, the two Friedmann equations are
\begin{align}
    3 M_p^2 H^2 &= \frac{F(\phi)}{2}\,\dot\phi^2 + V(\phi),
    \label{eq:Fr1} \\
    -2 M_p^2 \dot H &= F(\phi)\,\dot\phi^2,
    \label{eq:Fr2}
\end{align}
where $H \equiv \dot a / a$ is the Hubble parameter and dot stands for derivative with respect to the cosmic time. Moreover, varying the action with respect to $\phi$ gives the field equation of motion 
\begin{equation}\label{eq:eom}
    \ddot\phi + 3H\dot\phi + \frac{F_{,\phi}}{2F(\phi)}\,\dot\phi^2
    + \frac{V_{,\phi}}{F(\phi)} = 0,
\end{equation}
where $F_{,\phi} = dF/d\phi$ and $V_{,\phi} = dV/d\phi$. For $F = 1$, Eq.~\eqref{eq:eom} reduces to the standard Klein-Gordon equation. The term $\frac{F_{,\phi}}{2F}\,\dot\phi^2$ acts as an additional friction or anti-friction contribution in the scalar field dynamics. When $F_{,\phi} > 0$, it decelerates the field and modifies the effective slow-roll parameters, which in turn shifts the CMB predictions relative to the canonical case.


The inflationary dynamics is conveniently described through the slow-roll parameters, defined as the hierarchy
\begin{equation}
    \epsilon_1 \equiv -\frac{\dot H}{H^2}, \qquad
    \epsilon_{n+1} \equiv \frac{\dot\epsilon_n}{H\,\epsilon_n},
    \quad n \geq 1,
    \label{eq:srp}
\end{equation}
which also known as the Hubble slow-roll parameters. Another slow-roll parameter that is widely used in the literature is defined as
\begin{equation}\label{eq:eta_sr}
    \eta \equiv \frac{\ddot\phi}{H\dot\phi},
\end{equation}
which measures the rate of the field velocity in a Hubble time. During inflation, all these parameters are required to remain much smaller than unity, i.e., $\epsilon_n \ll 1$, which is the condition that defines the slow-roll regimes and ensures a successful inflation which last long enough. The first slow-roll parameter $\epsilon_1$ measures the rate of the Hubble parameter during a Hubble time. Smallness of this parameter indicates that that the expansion is quasi-de Sitter. Using Eq.~\eqref{eq:Fr2}, it can be rewritten as 
\begin{equation}\label{eq:eps1_exact}
    \epsilon_1 = \frac{3 F(\phi)\,\dot\phi^2}{F(\phi)\,\dot\phi^2 + V(\phi)},
\end{equation}
his expression shows explicitly how the coupling $F(\phi)$ enters the first slow-roll parameter. Moreover, when $\epsilon_1 \ll 1$ it is realized that at the time that $\epsilon_1 \ll 1$, the kinetic term is negligible compared to the potential, indicating that the potential energy dominates during the slow-roll inflation. Inflation proceeds while $\epsilon_1 < 1$ and ends at $\epsilon_1 = 1$. The slow-roll condition $|\eta| \ll 1$ means that the field acceleration $\ddot\phi$ is negligible compared to the Hubble drag term $H\dot\phi$. This parameter is related to the Hubble slow-roll parameters through
\begin{equation}\label{eq:eta_relation}
    \eta = \epsilon_1 - \frac{\epsilon_2}{2},
\end{equation}
and it reduces to $\eta \simeq \epsilon_1$ at leading slow-roll order.

Applying the slow-roll approximation, $F(\phi)\,\dot\phi^2 \ll V(\phi)$ and $\ddot\phi \ll H \dot\phi$, simplify the background equations~\eqref{eq:Fr1} and \eqref{eq:eom} so that
\begin{equation}\label{eq:sr_H}
    3M_p^2 H^2 \simeq V(\phi),
\end{equation}
\begin{equation}\label{eq:sr_phi}
    3H F(\phi)\,\dot\phi \simeq -V_{,\phi}.
\end{equation}

In most literature, the slow-roll parameters are defined in terms of the potential. So, it is also useful to defined the potential slow-roll parameters. Using the simplified dynamical equations \eqref{eq:sr_H} and \eqref{eq:sr_phi}, the two well know potential slow-roll parameters are given as~\cite{Barenboim:2007ii, Franche:2010yj}
\begin{equation}\label{eq:srp_V}
    \epsilon_V \equiv \frac{M_p^2}{2F(\phi)}
    \left(\frac{V_{,\phi}}{V}\right)^2, \qquad
    \eta_V \equiv \frac{M_p^2}{F(\phi)}\,\frac{V_{,\phi\phi}}{V}.
\end{equation}
One can see that for $F(\phi)$ these expressions reduce to the standard potential slow-roll parameters, while a non-trivial coupling $F(\phi)$ rescales both $\epsilon_V$ and $\eta_V$ relative to the canonical case. 

It should be noted that in the slow-roll regime, we have $\epsilon_1 \simeq \epsilon_V$. The scalar spectral index and tensor-to-scalar ratio at leading slow-roll order are given as 
\begin{equation}\label{eq:ns_r_sr}
    n_s \simeq 1 + 2\eta_V - 6\epsilon_V - \xi_V, \qquad
    r \simeq 16 \epsilon_V.
\end{equation}
where 
\begin{equation}
    \xi_V = \frac{F_{,\phi}}{F} \sqrt{2 F \epsilon_V}
\end{equation}
and the running of the scalar spectral index is defined as the logarithmic derivative of $n_s$ with respect to the wavenumber $k$, evaluated at the pivot scale,
\begin{equation}\label{eq:running}
    \alpha_s \equiv \frac{dn_s}{d\ln k}\bigg|_{k=k_{\rm pivot}}.
\end{equation}
Due to the appearance of the coupling term $F(\phi)$ in the denominator of $\epsilon_V$ and $\eta_V$, it rescales the parameters relative to the canonical prediction for the same potential. A suitable coupling term could suppress $\epsilon_V$ and $\eta_V$ during the inflationary phase, pushing $n_s$ toward unity. It is worth noting that the last term in the expression for $n_s$, namely $\xi_V$, is a direct consequence of the non-canonical kinetic coupling $F(\phi)$ and has no counterpart in the standard canonical inflation. This term vanishes identically when $F_{,\phi} = 0$, which includes the canonical limit $F = 1$, and the standard expression $n_s \simeq 1 + 2\tilde{\eta}_V - 6\tilde{\epsilon}_V$ is recovered, where $\tilde{\epsilon}_V = F(\phi) \epsilon_V$ and $\tilde{\eta}_V = F(\phi) \eta_V$. For a non-trivial coupling $F(\phi)$, this additional term introduces a further correction to the spectral index whose sign and magnitude depend on the form of $F(\phi)$ and its derivative $F_{,\phi}$. This is the essential physical mechanism by which the k-essence model under consideration here can restore the observational viability of the Starobinsky potential under the ACT constraints.

\subsection{Equations in terms of e-folds}
\label{subsec:efolds}
The total inflationary expansion is measured by the number of e-folds, defined as
\begin{equation}
    N \equiv \int_{t_i}^{t_e} H\,dt = \ln\frac{a_e}{a_i},
    \label{eq:efolds}
\end{equation}
where the subscript ${}_i$ and ${_e}$ stand to indicate the beginning and end of inflation, respectively. To successfully solve the horizon and flatness problems, it is required that inflation last roughly around $N \sim 55$--$65$ e-folds~\cite{Baumann:2009ds, Guth:1980zm, Linde:1981mu}.

To obtain the inflationary predictions accurately, we work with the exact background equations~\eqref{eq:Fr1} and \eqref{eq:eom}, and solve them numerically. It is therefore more practical to express these equations in terms of the number of e-folds $N$ rather than cosmic time, as this choice simplifies the numerical integration and makes it more convenient to connect the prediction to observation because all CMB-relevant modes cross the horizon within a narrow range of $N$. Using $d/dt = H\,d/dN$, the background dynamical equations \eqref{eq:Fr1}--\eqref{eq:eom} can be written as
\begin{align}
    3M_p^2 H^2 &= \frac{F(\phi)}{2}\,H^2\phi'^2 + V(\phi),
    \label{eq:Fr1_N} \\
    2M_p^2 H' &= -F(\phi)\,H \phi'^2,
    \label{eq:Fr2_N} \\
    \phi'' &= - \left(3 - \epsilon_1\right)\phi'
    - \frac{F_{,\phi}}{2F}\,\phi'^2
    - \frac{V_{,\phi}}{H^2 F},
    \label{eq:eom_N}
\end{align}
where 'prime' here indicates a derivative with respect to the e-fold $N$. With this change of variable, the slow-roll parameter could be read as $\epsilon_1 = -H'/H = F(\phi) \phi'^2 / 2$. This reformulation in terms of $N$, the starting point for the numerical integration and finding the observable quantities at the CMB scales is in the next section. 

In what follows, we specify the potential and coupling function, present the inflationary predictions from both the slow-roll approximation and a full numerical solution of Eqs.~\eqref{eq:Fr1_N}--\eqref{eq:eom_N}, and compare the resulting $(n_s, r)$ values against the ACT DR6 observational contours. We also analyze the attractor behaviour of the inflationary trajectory to confirm that the predictions are insensitive to the choice of initial conditions.

\section{Results}
\label{sec:results}

In this section, we present the main results of the introduced model. We first solve the background equations derived in Sec.~\ref{sec:kinfaltion} numerically and obtain the inflationary predictions for the scalar spectral index $n_s$, its running, and the tensor-to-scalar ratio $r$. The resulting $(n_s, r)$ values are then placed on the ACT DR6 observational contours to explore the consistency of the model with the most current available CMB data. We then examine the attractor behaviour of the inflationary solution through a phase-space analysis, and demonstrate that the slow-roll solution acts as an attractor for a wide range of initial conditions. 

The potential is taken as a Starobinsky-like form, Eq.\eqref{eq:starobinsky}~\cite{starobinsky:1980te}. 
This choice is not arbitrary, but it is motivated by theoretical considerations. The potential originates from the $R^2$ gravity action, appears in no-scale supergravity constructions~\cite{Ellis:2025zrf}, and corresponds to the $\alpha = 1$ member of the $\alpha$-attractor family~\cite{Kallosh:2013yoa}. It was in excellent agreement with CMB data before ACT DR6, and here we address the question of whether this theoretically compelling potential can also satisfy ACT DR6 constraints within the k-essence framework. The coupling function is chosen to be a power-law form $ F(\phi) = 1 + A\,\phi^n$,
where $V_0$ is the potential amplitude is fixed by the CMB normalisation and $A$ and $n$ are the free parameters of the coupling function that control the strength and field dependence of the kinetic modification.

Before presenting the results, we briefly describe the observational datasets used to confirm the viability of the model. The constraints on $n_s$ and $r$ come from three CMB dataset combinations, which have been summarized in Table~\ref{tab:datasets}. The first dataset is Planck 2018 combined with BICEP/Keck 2018 (BK18), which gives $n_s = 0.9649 \pm 0.0042$ (68\% CL) and $r < 0.036$ (95\% CL)~\cite{Planck:2018jri, BICEP:2021xfz}. The second dataset, which also can be addressed as the primary motivation for the this work, is the joint combination of Planck, ACT DR6, DESI BAO, and BK18 (P-ACT-LB-BK18), which shifts the spectral index significantly upward to $n_s = 0.9743 \pm 0.0034$ (68\% CL) with $r < 0.038$ (95\% CL)~\cite{ACT:2025fju, ACT:2025tim}. Due to this upward shift in the scalar index, the Starobinsky potential in the canonical framework is currently in tension with the data. The third dataset is the combination of SPT-3G, Planck, and ACT (CMB-SPA), which gives $n_s = 0.9684 \pm 0.0030$ (68\% CL)~\cite{SPT-3G:2025bzu}, intermediate between the two previous values. This data does not provide a constraint on the tensor-to-scalar ratio $r$, and here we only use it to compare the scalar spectral index \cite{Ellis:2025zrf}. 
\begin{table}[t]
\caption{The table presents the summary of the three observational datasets we use in this work to confirm the validity of the model. The datasets put constraints on the scalar spectral index $n_s$ and the tensor-to-scalar ratio $r$ at the pivot 
scale $k_* = 0.05~{\rm Mpc^{-1}}$.}
\label{tab:datasets}
\centering
\setlength{\tabcolsep}{6pt}
\begin{tabular}{@{}lcc@{}}
\toprule
Dataset & $n_s$ & $r$ \\
\midrule
Planck 2018~\cite{Planck:2018jri}
    & $0.9649 \pm 0.0042$ & $< 0.10$ \\
Planck 2018 + BK18~\cite{BICEP:2021xfz}
    & --- & $< 0.036$ \\
P-ACT-LB~\cite{ACT:2025fju}
    & $0.9743 \pm 0.0034$ & --- \\
P-ACT-LB-BK18~\cite{ACT:2025tim}
    & --- & $< 0.038$ \\
CMB-SPA~\cite{SPT-3G:2025bzu}
    & $0.9684 \pm 0.0030$ & --- \\
\bottomrule
\end{tabular}
\end{table}

\subsection{Inflationary predictions and comparison with ACT DR6}
\label{subsec:rns}

The free parameters of the model are the amplitude $V_0$, the coupling amplitude$A$, and the power index $n$. The first parameter is set by the CMB scalar power spectrum, and it does not influence the scalar spectral index, its running, or the tensor-to-scalar ratio. We therefore focus on the effect of $A$ and $n$ on these observables and identify the parameter values for which the predictions fall within the ACT DR6 observational contours. 

The procedure for obtaining the prediction is as follows. We solve the background equations~\eqref{eq:eom_N} numerically by setting initial conditions at the beginning of inflation, $N = 0$, where the slow-roll approximation is still valid. For each choice of $(A, n)$, we evolve the equations and keep only those parameter sets for which the inflationary phase lasts at least $N = 60$ e-folds. The inflationary observables $n_s$, $\alpha_s$, and $r$ are then estimated at $N = 55$ and $N = 60$ e-folds before the end of inflation, corresponding to the moment of horizon crossing of the pivot scale, $k_{\rm pivot} = 0.05 \; {\rm Mpc^{-1}}$, which is directly linked to CMB observations. To gain a better understanding of how each parameter affects the predictions, we study the model in two complementary ways. First by varying $A$ while keeping the rest fixed, and then by varying the power index $n$. Then, for each case, we project the results on $r-n_s$ and $\alpha_s -n_s$ counters of ACT DR6. 

\begin{figure*}[t]
    \centering
    \subfigure{\includegraphics[width=0.45\columnwidth]{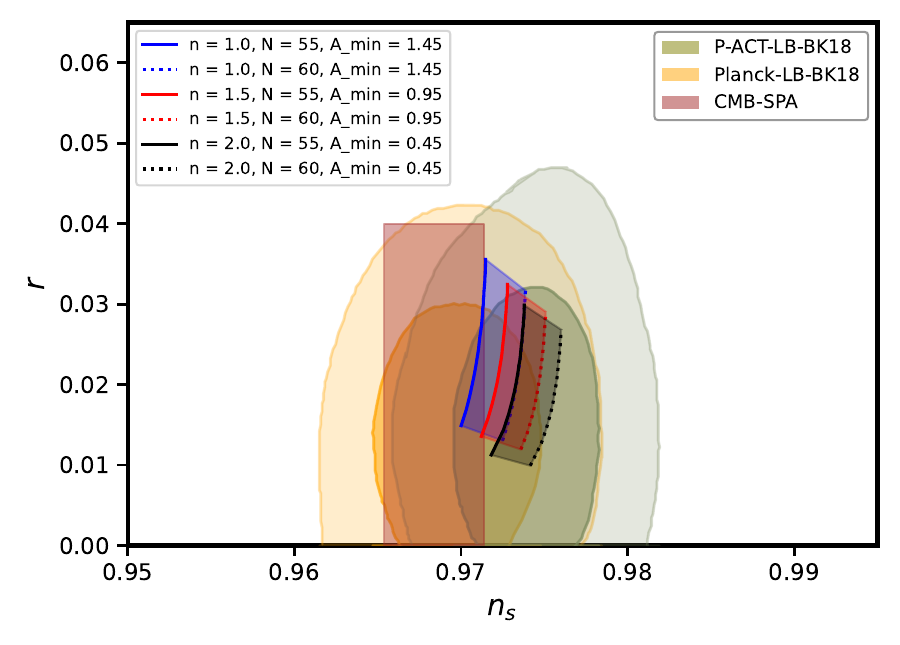}}
    \subfigure{\includegraphics[width=0.45\columnwidth]{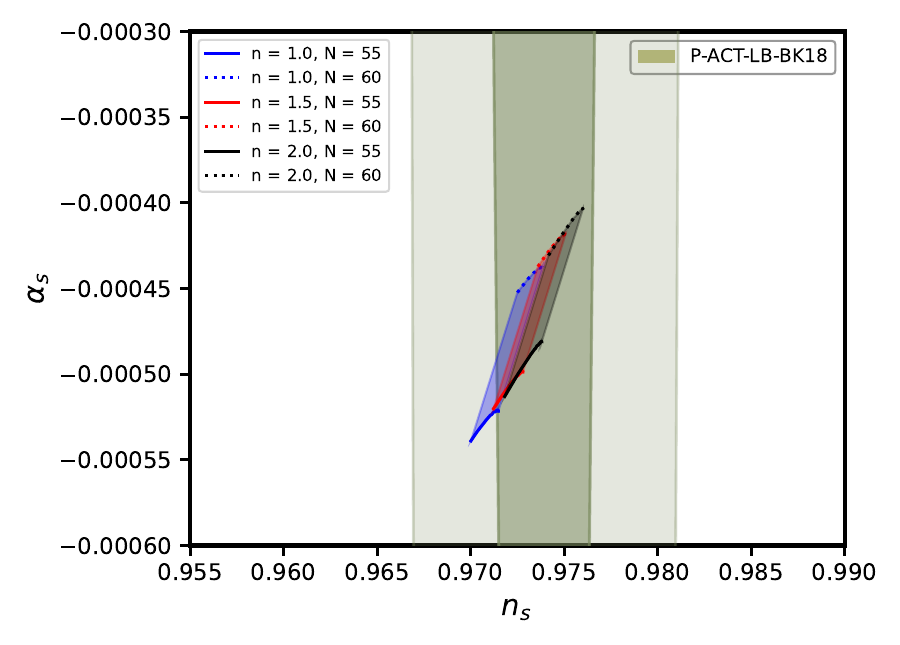}}
    \caption{right panel and left panel show the $r$--$n_s$ and $\alpha_s$--$n_s$ results, respectively, for three values of the power index, $n = 1.0$ (blue), $1.5$ (red), and $2.0$ (black), while the coupling amplitude $A$ varies. For each set, the filled region shows the resulting $(n_s, r)$ for e-folds in the range $N = 55$--$60$, with the boundaries corresponding to $N = 55$ (solid lines) and $N = 60$ (dotted line). The model predictions lie within the $1\sigma$ region of the ACT DR6 joint analysis (P-ACT-LB-BK18)~\cite{ACT:2025fju, ACT:2025tim}.}
    \label{fig:rns_A}
\end{figure*}
\begin{figure*}
    \centering
    \includegraphics[width=0.8\linewidth]{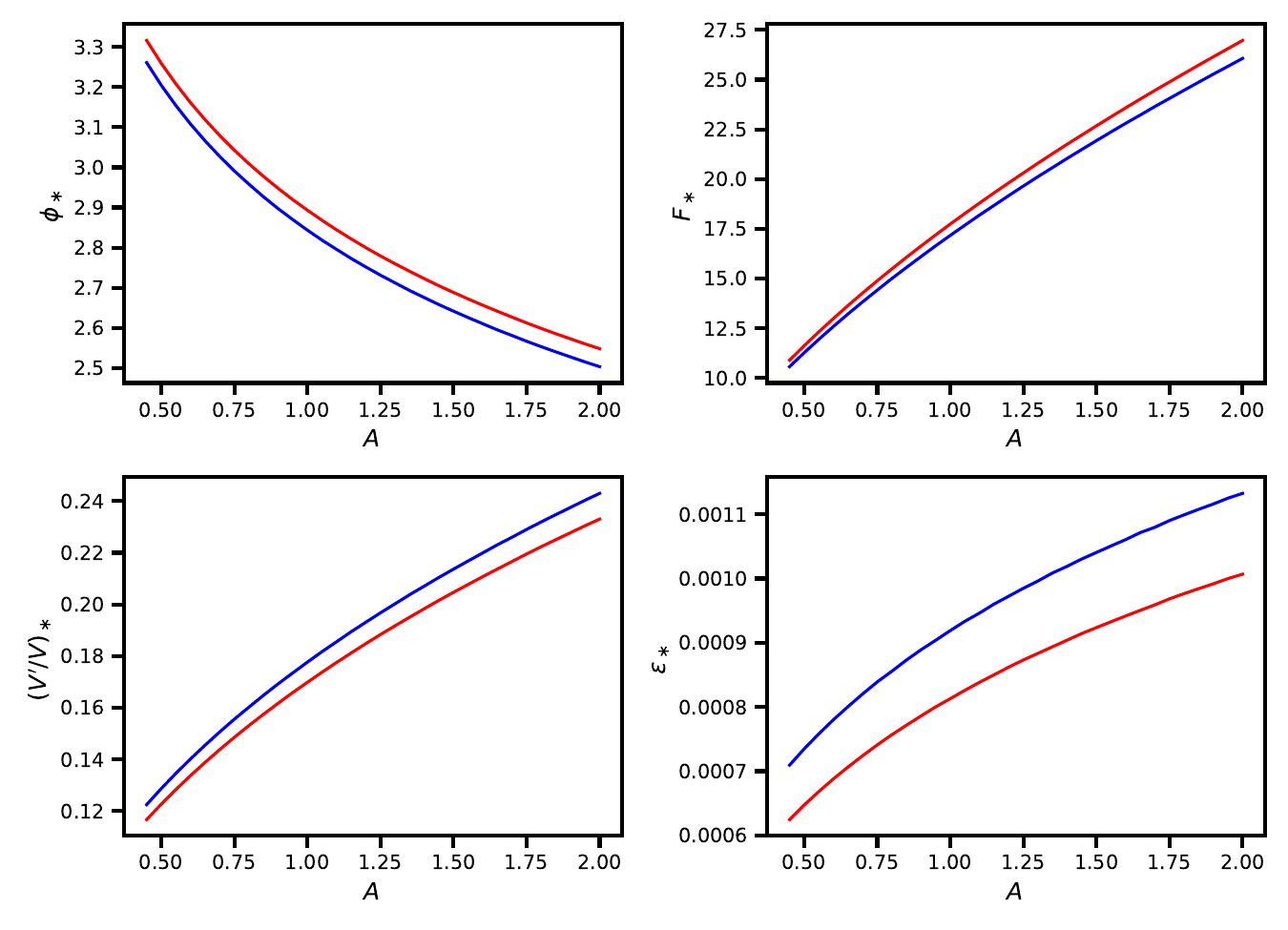}
    \caption{The behavior of the a)scalar field, b)coupling function, c) $V'/V$ ratio, and d) slow-roll parameters versus the coupling parameter $A$ are shown at the crossing time, where $n = 2.0$. The red and blue lines are related to the crossing e-folds $N = 55$ and $N = 60$ e-folds before inflation. }
    \label{fig:checking}
\end{figure*}

As the first case, we consider fixed values of the power index $n = 1.0$, $1.5$, and $2.0$, and vary the coupling amplitude from a minimum value up to $A = 15.5$. The minimum value of $A$ is set separately for each $n$ such that the inflationary phase lasts at least $60$ e-folds. 
The resulting $r$--$n_s$ curves for $N = 55$ (solid lines) and $N = 60$ (dotted lines) are shown in the left panel of Fig.~\ref{fig:rns_A}, projected on ACT DR6 observational contours. The results fall within the $1\sigma$ region, confirming the consistency of the model with ACT DR6 and demonstrating that a Starobinsky-like potential, which is disfavoured in the canonical framework, becomes consistent with data once embedded in non-canonical settings. As $A$ increases, in general, a lower $r$ is expected, because a higher $A$ leads to higher coupling. However, this is not the whole story. One should also note that, by increasing $A$, the total amount of inflationary e-folds also increases so that the field value at horizon crossing, $\phi_\star$, decreases to maintain a fixed $N = 55$ or $N = 60$ before the end of inflation. In other words, a smaller field is required to generate the same amount of e-folds as $A$ increases. This smaller one $\phi_\star$ in turn affects the values of the coupling, potential, and derivative of potential at the horizon crossing. Fig.\ref{fig:checking} shows the behaviour of the scalar field, coupling $F$, the ratio $V_{,\phi}/V$, and the slow-roll parameter $\epsilon_V$ at the crossing time versus the amplitude parameter $A$. It can be clearly seen that by increasing $A$, the scalar field at the horizon crossing decreases to maintain crossing time at $N = 55 (65)$ before the end of inflation. Meanwhile, the coupling at the crossing time increases in general, however the ratio $V_{,\phi}/V$ decreases. Gathering all together, the slow-roll parameter $\epsilon_V$ increases with $A$. 
Considering all these effects together on the slow-roll parameters $\epsilon_V$, $\eta_V$, and $\xi_V$, it is found that both $n_s$ and $r$ increase with $A$, while the result remains within the $1\sigma$ region of ACT DR6. Moreover, we observe the dependency of the number of e-folds so that it follows the standard pattern. Higher spectral index and lower tensor-to-scalar ratio are obtained with increasing e-folds. The running of the spectral index $\alpha_s$ is shown in the right panel of Fig.~\ref{fig:rns_A}. For all cases, the running of the spectral index is negative; however, it remains consistent with the ACT DR6 constraints across the full range of $A$ considered, in which it stands in $1\sigma$ region for both crossing e-folds $N = 55$ and $60$. Higher e-folds, as expected, lead to higher scalar spectral index and lower running, which is illustrated in the figure. 

\begin{figure*}[t]
    \centering
    \subfigure{\includegraphics[width=0.45\columnwidth]{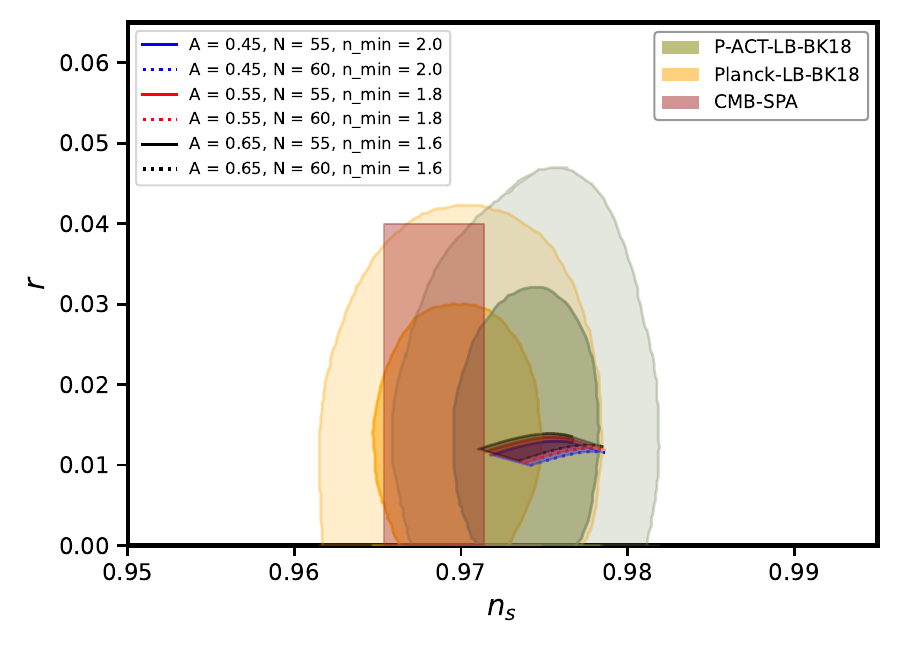}}
    \subfigure{\includegraphics[width=0.45\columnwidth]{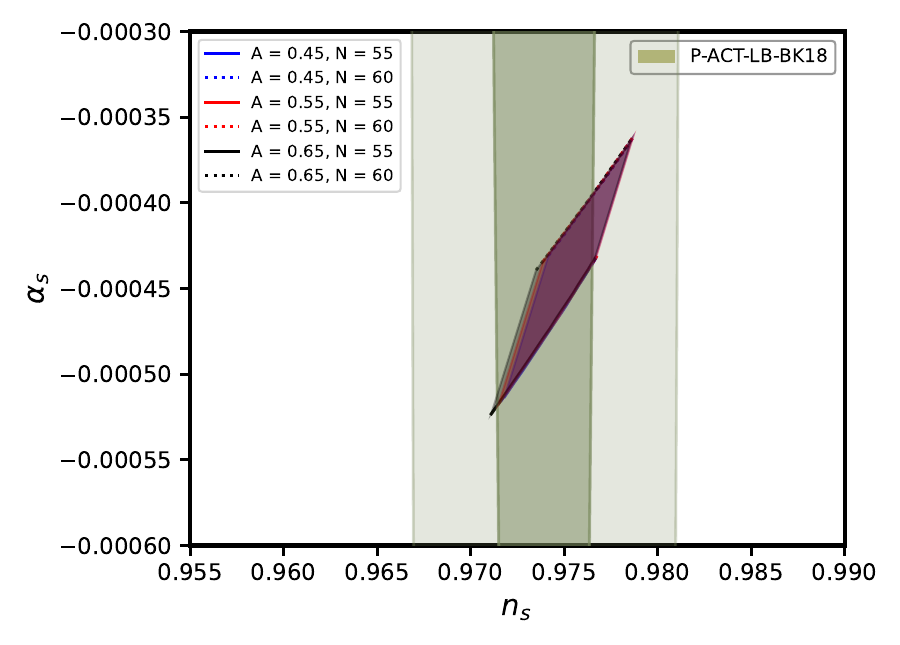}}
    \caption{right panel and left panel show the $r$--$n_s$ and $\alpha_s$--$n_s$ results, respectively, for three values of the coupling amplitude parameter, $A = 0.45$ (blue), $0.55$ (red), and $0.65$ (black), while the power index $n$ varies. For each set, the filled region shows the resulting $(n_s, r)$ for e-folds in the range $N = 55$--$60$, with the boundaries corresponding to $N = 55$ (solid lines) and $N = 60$ (dotted line). The model predictions lie within the $1\sigma$ region of the ACT DR6 joint analysis (P-ACT-LB-BK18)~\cite{ACT:2025fju, ACT:2025tim}.}
    \label{fig:rns_q}
\end{figure*}

In the second case, the amplitude of the coupling is fixed to $A = 0.45$, $0.55$, and $0.65$, and the power index varies from a minimum value up to $n = 5.4$. The minimum value of $n$ is set separately for each $n$ such that the inflationary phase lasts at least $60$ e-folds. The resulting $r$--$n_s$ curves for $N = 55$ (solid lines) and $N = 60$ (dotted lines) are displayed in the left panel of Fig.~\ref{fig:rns_q}, where they are projected over the ACT observational contour. We observe almost the same behaviour as the first case. At first sight, one might expect that increasing $n$ leads to a larger coupling $F(\phi)$, which would suppress it, $\epsilon_V$, and therefore, a lower tensor-to-scalar ratio is expected. However, this is not what is found. Increasing $n$ leads to an inflation with higher e-folds of expansion, which forces the scalar field at horizon crossing $\phi_\star$ to take a smaller value to maintain a fixed $N = 55$ or $N = 60$ before the end of inflation. This reduction, $\phi_\star$, together with an increasing power index, $n$ leads to higher effective coupling $F(\phi_\star)$; however, it leads to higher values for the ratio $V_{,\phi}/V$. As a result, $\epsilon_V$ increases with $n$, leading to a higher $r$. Compared to the first case, we realize less enhancement in $r$ and more increase in $n_s$. For both $N = 55$ and $N = 60$, the model predictions lie within the $1\sigma$ region of ACT DR6 over the whole range of $n$. This confirms that the k-essence framework with a Starobinsky-like potential has the flexibility to accommodate the new observational constraints through the kinetic coupling alone, without modifying the gravitational sector. The right panel of Fig.\ref{fig:rns_q}, shows the corresponding running of the spectral index projected on the ACT DR6 data. The resulting running is negative, and it can remain within the $1\sigma$ region.

Table~\ref{tab:results} presents the numerical values of the inflationary observables quantities $n_s$, $\alpha_s$, and $r$ for different selected values of the coupling parameters $A$ and $n$. The quantities are evaluated first by assuming that the horizon crossing times occurs at $N = 55$ e-folds before the end of inflation, and second by assuming it occurs at $N = 60$ e-folds before the end of inflation. For each set of parameters, the table reports two categories of the result. The first one corresponds to the results obtained by using the potential slow-roll parameters $\epsilon_V$ and $\eta_V$, defined in Eq.~\eqref{eq:srp_V}, together with the spectral expressions in Eq.~\eqref{eq:ns_r_sr}. The second category corresponds to the result where the Hubble slow-roll parameters $\epsilon_1$ and $\epsilon_2$, defined in Eq.~\eqref{eq:srp}, are used. These slow-roll parameters are estimated numerically, and the scalar spectral index and the tensor-to-scalar ratio are obtained as $n_s = 1 - (2\epsilon_1 + \epsilon_2)$ and $r = 16 \epsilon_1$. The differences between the results of these two categories are very small, confirming that the slow-roll approximation is reliable at the horizon crossing times. Such a deviation in general is expected because the potential slow-roll parameters are obtained under the slow-roll approximations, while the Hubble slow-roll parameters carry more accurate information. Then, a small error relative to the exact Hubble slow-roll parameters at the crossing times is expected, and in which for all cases, one observes that the potential slow-roll values are slightly smaller. Additionally, the results for all ranges of parameters agree perfectly with the observational data ACT DR6.
\begin{table*}[t]
\caption{The inflationary observables $n_s$, $r$, and $\alpha_s$ for 
different values of the coupling parameters $A$ and $n$, evaluated at 
$N = 55$ and $N = 60$ e-folds before the end of inflation. For each set of parameters, the table shows two different categories of $n_s$, $r$, and $\alpha_s$ in which the first is related to the ones estimated based on the potential slow-roll parameters, and the second, distinguished by subscript $^H$, is related to the observable parameters estimated based on the Hubble slow-roll parameters.}
\label{tab:results}
\centering
\setlength{\tabcolsep}{8pt}
\begin{tabular}{@{}cc|ccc|ccc@{}}
\toprule
 & & \multicolumn{3}{c|}{$N = 55$} & \multicolumn{3}{c}{$N = 55$} \\
\midrule
$n$ & $A$ & $n_s$ & $r$ & $\alpha_s$ & $n_s^{(H)}$ & $r^{(H)}$ & $\alpha_s^{(H)}$ \\
\midrule
$1.0$ & $1.5$ & $0.9701$ & $0.0152$ & $-5.38 \times 10^{-4}$ & $0.9704$ & $0.0151$ & $-5.27 \times 10^{-4}$ \\ 
$1.0$ & $3.5$ & $0.9708$ & $0.0216$ & $-5.27 \times 10^{-4}$ & $0.9711$ & $0.0214$ & $-5.17 \times 10^{-4}$  \\
$1.0$ & $5.5$ & $0.9711$ & $0.0256$ & $-5.24 \times 10^{-4}$ & $0.9714$ & $0.0253$ & $-5.14 \times 10^{-4}$  \\
$1.5$ & $1.5$ & $0.9716$ & $0.0162$ & $-5.13 \times 10^{-4}$ & $0.9719$ & $0.0160$ & $-5.03 \times 10^{-4}$ \\ 
$1.5$ & $3.5$ & $0.9723$ & $0.0215$ & $-5.04 \times 10^{-4}$ & $0.9725$ & $0.0214$ & $-4.95 \times 10^{-4}$  \\
$1.5$ & $5.5$ & $0.9725$ & $0.0248$ & $-5.01 \times 10^{-4}$ & $0.9728$ & $0.0246$ & $-4.92 \times 10^{-4}$  \\
$2.0$ & $1.5$ & $0.9729$ & $0.0167$ & $-4.94 \times 10^{-4}$ & $0.9732$ & $0.0165$ & $-4.84 \times 10^{-4}$  \\
$2.0$ & $3.5$ & $0.9734$ & $0.0212$ & $-4.86 \times 10^{-4}$ & $0.9736$ & $0.0210$ & $-4.77 \times 10^{-4}$  \\ 
$2.0$ & $5.5$ & $0.9736$ & $0.0238$ & $-4.84 \times 10^{-4}$ & $0.9738$ & $0.0236$ & $-4.75 \times 10^{-4}$ \\ 
\midrule
 & & \multicolumn{3}{c|}{$N = 60$} & \multicolumn{3}{c}{$N = 60$} \\
\hline
$1.0$ & $1.5$ & $0.9725$ & $0.0133$ & $-4.52 \times 10^{-4}$ & $0.9728$ & $0.0132$ & $-4.44 \times 10^{-4}$ \\
$1.0$ & $3.5$ & $0.9732$ & $0.0190$ & $-4.43 \times 10^{-4}$ & $0.9735$ & $0.0188$ & $-4.35 \times 10^{-4}$ \\
$1.0$ & $5.5$ & $0.9735$ & $0.0226$ & $-4.39 \times 10^{-4}$ & $0.9737$ & $0.0224$ & $-4.32 \times 10^{-4}$ \\
$1.5$ & $1.5$ & $0.9740$ & $0.0143$ & $-4.31 \times 10^{-4}$ & $0.9742$ & $0.0141$ & $-4.23 \times 10^{-4}$ \\
$1.5$ & $3.5$ & $0.9746$ & $0.0191$ & $-4.23 \times 10^{-4}$ & $0.9748$ & $0.0190$ & $-4.16 \times 10^{-4}$ \\
$1.5$ & $5.5$ & $0.9748$ & $0.0220$ & $-4.20 \times 10^{-4}$ & $0.9750$ & $0.0219$ & $-4.13 \times 10^{-4}$ \\
$2.0$ & $1.5$ & $0.9751$ & $0.0148$ & $-4.14 \times 10^{-4}$ & $0.9754$ & $0.0147$ & $-4.07 \times 10^{-4}$ \\
$2.0$ & $3.5$ & $0.9756$ & $0.0189$ & $-4.08 \times 10^{-4}$ & $0.9758$ & $0.0188$ & $-4.01 \times 10^{-4}$ \\
$2.0$ & $5.5$ & $0.9758$ & $0.0213$ & $-4.06 \times 10^{-4}$ & $0.9760$ & $0.0211$ & $-3.99 \times 10^{-4}$ \\

\bottomrule
\end{tabular}
\end{table*}

The two analyses presented above establish a clear result. A Starobinsky-like potential, which is disfavoured in the canonical single-field framework by the ACT DR6 data~\cite{ACT:2025fju, ACT:2025tim}, is brought into full agreement with observations through a simple power-law kinetic coupling, with no modification to the gravitational sector. This finding is particularly interesting, in which a well-motivated modification of the kinetic structure of the scalar field is sufficient to bring this plateau-type potential into agreement with the ACT DR6 data. The model can achieve this for a well-defined range of the coupling parameters so that the resulting $n_s$, $\alpha_s$, and $r$ all fall within the $1\sigma$ region of ACT DR6. Above, we have considered a power-law form of the coupling function; however, the k-inflation framework employed in this work also satisfies the ACT constraints for other classes of coupling functions, such as an exponential coupling. This demonstrates that our results are not tied to a specific choice of coupling function, but rather reflect the robustness of the underlying theoretical framework.

\subsection{Power spectrum}
\label{subsec:ps}
\begin{figure}
    \centering
    \includegraphics[width=0.5\linewidth]{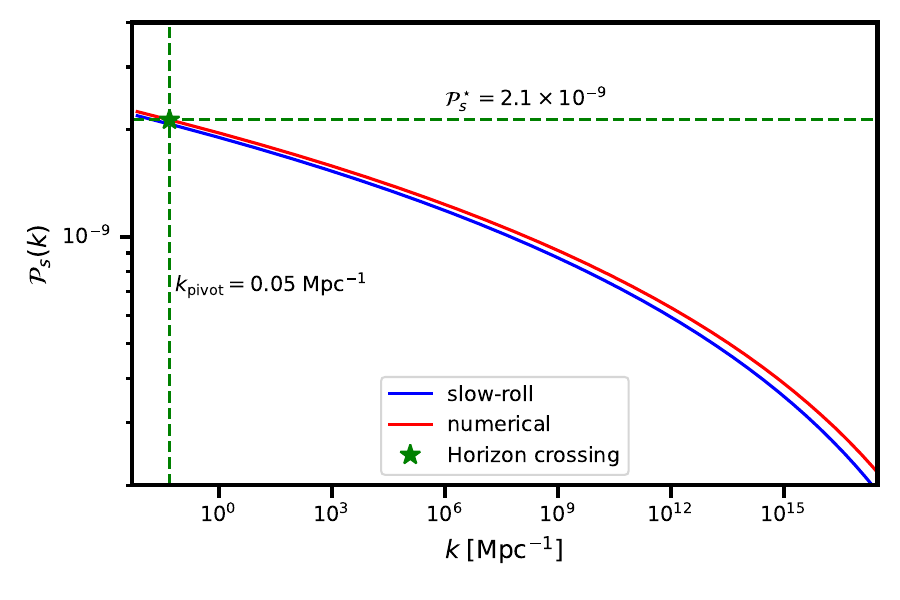}
    \caption{The figure shows the numerically obtained power spectrum (red line) by solving Mukhanov-Sasaki equation, against the analytical power spectrum (blue line). The parameters of the model are taken as $A = 0.55$, $q = 2.0$, and $m = 2.41$. The star stands for the horizon crossing time, where the obtained power spectrum is $\mathcal{P}_s^{\rm (num)}(k_{\rm pivot}) = 2.13 \times 10^{-9}$.}
    \label{fig:ps}
\end{figure}

\begin{figure}
    \centering
    \includegraphics[width=0.5\linewidth]{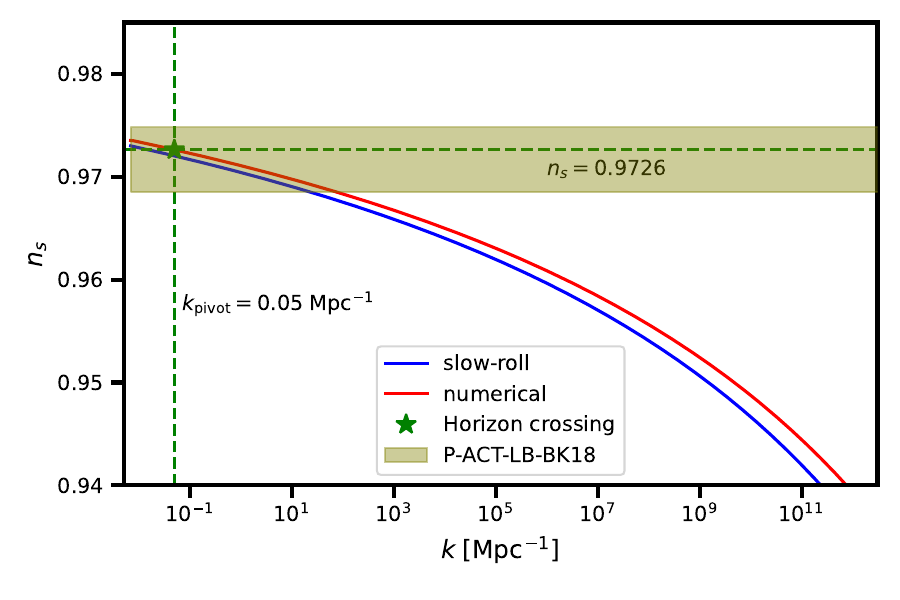}
    \caption{The figure shows the numerically obtained numerical scalar spectral index (red line) using definition \eqref{eq:ns_numerical}, and the blue line shows the analytical slow-roll estimate from Eq.~\eqref{eq:ns_r_sr}. The parameters of the model are taken as $A = 0.55$, $q = 2.0$, and $m = 2.41$. The green star stands for the horizon crossing time of the pivot scale $k_{\rm pivot} = 0.05~{\rm Mpc^{-1}}$, where the numerical and slow-roll estimates give $n_s = 0.9726$ and $n_s = 0.9720$, respectively.}
    \label{fig:ns_numerical}
\end{figure}

Scalar perturbations generated during inflation are described by the gauge-invariant comoving curvature perturbation $\zeta$. For the action~\eqref{eq:action}, its quadratic action reads~\cite{Garriga:1999vw,Chen:2006nt,Kobayashi:2010cm}
\begin{equation}\label{S2}
    S^{(2)} = \int dt\,d^3x\,a^3\,\frac{F(\phi)\dot\phi^2}{H^2}\left[\dot\zeta^2 - \frac{c_s^2}{a^2}(\nabla\zeta)^2\right].
\end{equation}
It should be noted that the propagation of the curvature perturbations is $c_s = 1$ because the action is linear in $X$ \cite{Armendariz-Picon:1999hyi, Garriga:1999vw}. To keep the coefficient of the kinetic term positive, it is required that the coupling function satisfy the condition $F(\phi) > 0$, which can be guaranteed by a proper choice of the coupling function. Defining the variable $v_k = z\,\zeta_k$ with $z = a\dot\phi\sqrt{F(\phi)}/H$, the Mukhanov-Sasaki equation is read as \cite{Garriga:1999vw,Kobayashi:2010cm}
\begin{equation}\label{MS}
    v_k'' + \left(k^2 - \frac{z''}{z}\right)v_k = 0,
\end{equation}
where a prime indicates a derivative with respect to the conformal time $\tau$ defined as $a d\tau = dt$. Deep inside the horizon, $k^2 \gg |z''/z|$, the evolution of the modes are well approximated by free oscillators, and the initial state is well described by the Bunch-Davies vacuum state \cite{Baumann:2009ds},
\begin{equation}\label{BD}
    v_k(\tau) \xrightarrow{k|\tau|\to\infty} \frac{e^{-ik\tau}}{\sqrt{2k}}.
\end{equation}
Then, the mode exits the horizon, and after some time it will be deep outside the horizon, $k^2 \ll |z''/z|$, where the curvature perturbations freeze. The scalar power spectrum is then computed by 
\begin{equation}\label{Ps}
    \mathcal{P}_\zeta(k) = \frac{k^3}{2\pi^2}\,|\zeta_k|^2\bigg|_{k\ll aH}.
\end{equation}
Applying the slow-roll approximations, the equations are simplified in which the scalar power spectrum are estimated analytically by 
\begin{equation}\label{Ps_sr}
    \mathcal{P}_\zeta^{\rm sr}(k) \simeq \frac{H^2}{8\pi^2 M_p^2 \epsilon_1}\bigg|_{k=aH},
\end{equation}
Considering Eq.\eqref{BD} as the initial condition, and using the background solution obtained in the previous subsection, the Mukhanov-Sasaki equation~\eqref{MS} is solved numerically for each mode $k$, and the resulting scalar power spectrum is shown in Fig.~\ref{fig:ps}. TThe red line stand for the full numerical result obtained from the definition~\eqref{Ps}, while the blue line shows the analytical slow-roll estimate from Eq.~\eqref{Ps_sr}. It is observed that differences are very small near the horizon crossing time showing that the the slow-roll conditions are well satisfied and the approximation is reliable. The green star show the horizon crossing time where the pivot mode $k_{\rm pivot} = 0.05 \; {\rm Mpc^{-1}}$ cross the horizon. the corresponding power spectrum is $\mathcal{P}_s^\star = 2.13 \times 10^{-9}$ which is consistent with data. However, as approaching to the end of inflation, the two power spectrum deviate from each other and the difference grows. This is because by approaching the end of inflation the slow-roll approximation is weakened and finally it is violated, as at the end of inflation $\epsilon_1 = 1$. The numerical result capture this behavior correctly, as it obtained by solving the complete evolution equations and it is valid for the whole inflationary phase, including near its end. This comparison confirms that at the CMB scale, the slow-roll approximation is valid and the obtained results are reliable. However, the full numerical approach is necessary to capture the accurate description at smaller scales. 
The scalar spectral index can also be extracted directly from the numerical power spectrum through the general definition
\begin{equation}\label{eq:ns_numerical}
    n_s - 1 = \frac{d\ln\mathcal{P}_s}{d\ln k}\bigg|_{k = k_{\rm pivot}},
\end{equation}
presented in Fig.\ref{fig:ns_numerical} against the slow-roll approximated scalar spectral index obtained from Eq.\eqref{eq:ns_r_sr}. The result is found to be in close agreement with the slow-roll estimate of Eq.~\eqref{eq:ns_r_sr}, providing a direct validation of the slow-roll approximation at the CMB pivot scale for the parameter values considered in this work.

\subsection{Attractor behavior}
\label{subsec:attractor}

It is crucially important that any viable inflationary model is not sensitive to the choice of the initial conditions. In general, if the inflationary trajectory is an attractor in the phase space of the scalar field, then different initial values of $\dot\phi$ at the same initial field value $\phi_0$ converge rapidly to the same slow-roll solution. Here, we are going to explore whether this feature holds for the k-essence model defined by action~\eqref{eq:action}.

\subsubsection{Hamilton-Jacobi formulation}
\label{subsubsec:HJ}

The Hamilton-Jacobi formalism provides a clean analytical framework to study the attractor behaviour. In this formalism, the Hubble parameter $H$ is treated as a function of the scalar field rather than of time. By changing the variable and working with scalar field $\phi$ instead  of time $t$, one has $\dot{H} = \dot\phi H_{,\phi}$, and the equation \eqref{eq:Fr2}, can be read as
\begin{equation}\label{eq:HJ_phidot}
    \dot\phi = -\frac{2M_p^2}{F(\phi)}\,H'(\phi),
\end{equation}
By substituting this into Eq.~\eqref{eq:Fr1}, one reaches the Hamilton-Jacobi equation for the k-essence model as
\begin{equation}\label{eq:HJ}
    \frac{2M_p^4}{F(\phi)}\left[H'(\phi)\right]^2
    - 3M_p^2 H^2(\phi) = -V(\phi).
\end{equation}
The advantage of the above equation is that one can consider how small perturbations around any given solution $H_0(\phi)$ evolve. In this regard, we consider a small perturbation over the background solution as $H(\phi) = H_0(\phi) + \delta H(\phi)$ where $\delta H \ll H_0$. Substituting into Eq.~\eqref{eq:HJ} and keeping up to the linear order of the perturbation, leads to the following expression 
\begin{equation}\label{eq:HJ_perturb}
    \frac{4M_p^4}{F(\phi)}\,H_0'(\phi)\,\delta H'(\phi)
    = 6M_p^2 H_0(\phi)\,\delta H(\phi),
\end{equation}
which has a general solution
\begin{equation}\label{eq:deltaH}
    \delta H(\phi) = \delta H(\phi_i) \exp\!\left[
    \frac{3}{2M_p^2}\int_{\phi_i}^{\phi}
    \frac{F(\phi) H_0(\phi)}{H_0'(\phi)}\,d\phi'
    \right].
\end{equation}
For the case of the Starobinsky potential, the terms $H_0'$ and $d\phi$ have opposite signs as the field rolls toward the minimum of the potential. Since both the Hubble parameter $H$ and the coupling $F$ are positive, the integrand in the exponent is negative definite throughout the inflationary phase. This leads to the conclusion that the perturbation $\delta H$ decays exponentially as $\phi$ rolls down. Therefore, any solution to Eq.~\eqref{eq:HJ} that supports inflation is an attractor, meaning that trajectories starting from different initial field velocities converge to the inflationary solution as the field evolves~\cite{Liddle:1994dx,Kinney:1997ne,Guo:2003nz}. The coupling $F(\phi)$, which appears in the exponent, changes the rate of convergence relative to the canonical case; however, the attractor behaviour is preserved since the coupling is positive.

\subsubsection{Phase-space analysis}
\label{subsubsec:phase}

\begin{figure}[t]
    \centering
    \includegraphics[width=0.5\linewidth]{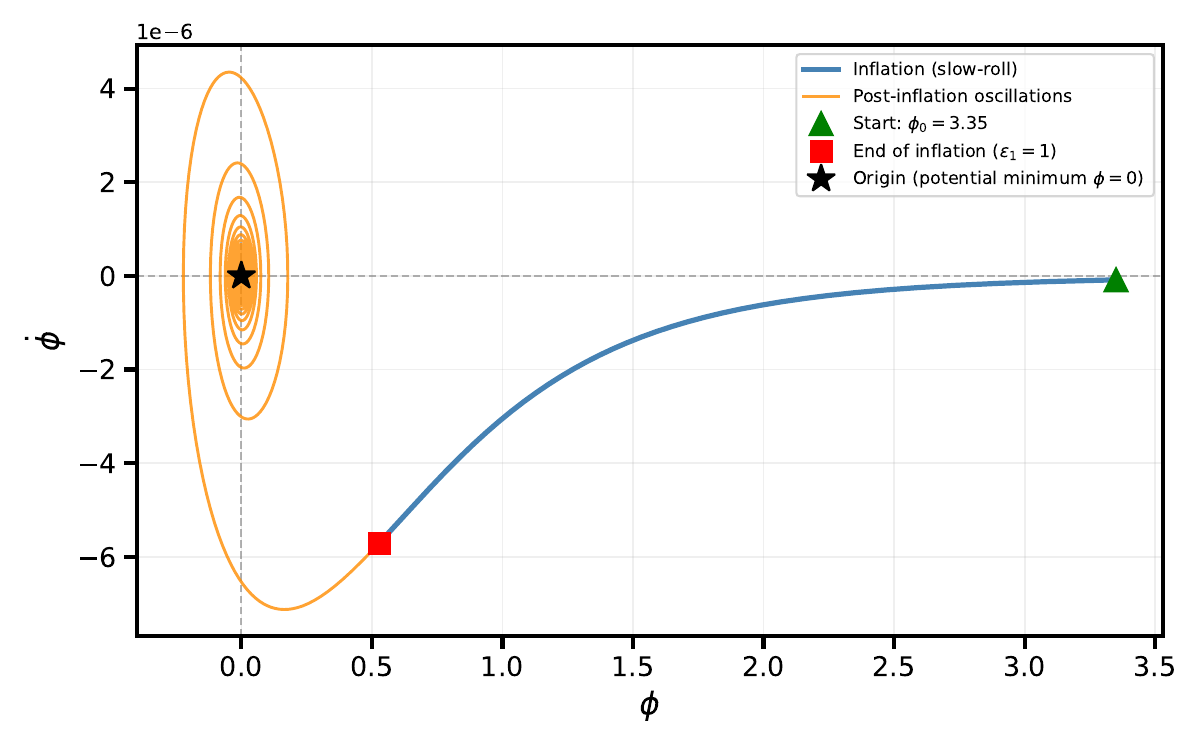}
    \caption{The trajectory of the scalar field in the $(\phi, \dot\phi)$ phase plane for a representative set of model parameters as $n = 2.0$ and $A = 0.55$. The trajectory begins at large field, shown by a green triangle. The field slowly rolls down the potential where we have inflationary phase, shown by blue line, and then it end at the red spot. After inflation, the field continues rolling downs toward the minimum of the potential, and starts oscillating around the minimum. it is shown by the orange lines where the shrinking spiral motion on phase space shows the oscillation of the field around the minimum.}
    \label{fig:phase_single}
\end{figure}

To verify the attractor behaviour of the solution numerically and also to provide a visualisation of the structure of the phase space, we work directly with the original background equations \eqref{eq:Fr1}-\eqref{eq:eom}. From these equations, we have
\begin{align}
    \frac{d\phi}{dt} &= \dot\phi,
    \label{eq:phase1} \\
    \frac{d\dot\phi}{dt} &= -3H\dot\phi
    - \frac{F_{,\phi}}{2F(\phi)}\,\dot\phi^2
    - \frac{V_{,\phi}}{F(\phi)},
    \label{eq:phase2}
\end{align}
in which the Hubble parameter, which $H$ should be determined at each point through the Friedmann equation \eqref{eq:Fr1}. By setting an initial condition $(\phi_0, \dot\phi_0)$, the system can be integrated numerically, giving a trajectory in the $(\phi, \dot\phi)$ phase plane. Figure~\ref{fig:phase_single} shows one representative trajectory for a chosen set of parameters of the model.  The curve starts from a large initial field value with a specific initial velocity $\dot\phi_0$, as shown by a green triangle. During inflation, the scalar field rolls down the potential, and it rolls toward the smaller values of $\phi$. As $\epsilon_1$ approaches unity, inflation ends, determined by a red square. The trajectory during inflation is shown by the solid blue line. Then, the field enters the post-inflationary phase, where the scalar field oscillates around the minimum of the potential. This phase is exhibited by a shrinking spiral pattern around the origin in the $(\phi, \dot\phi)$ plane, where it loses energy due to the friction terms. This spiral motion toward the origin is what one expects from an inflaton oscillating around the minimum of its potential after inflation ends. It confirms that the dynamical system is well-behaved beyond the inflationary epoch.

\begin{figure}[t]
    \centering
    \includegraphics[width=0.5\linewidth]{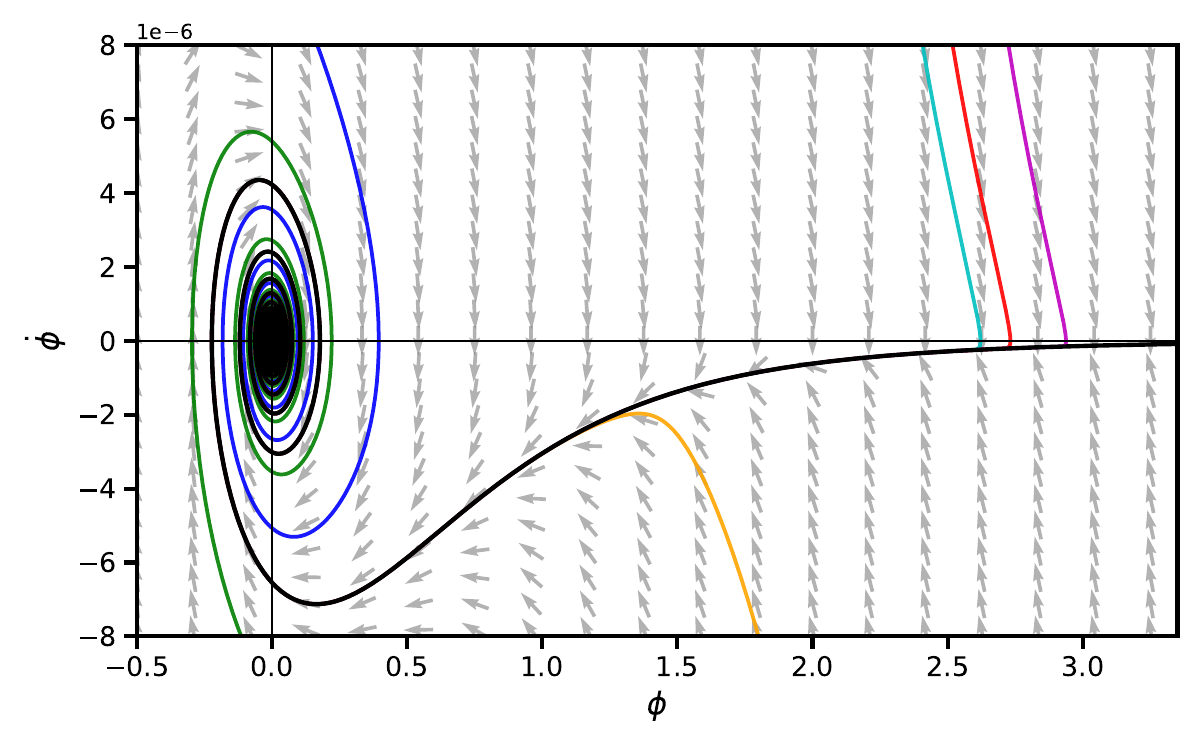}
    \caption{Phase portrait of the k-essence inflation model. Multiple trajectories (colored curves) start from the same initial field value $\phi_0$ with different initial velocities $\dot\phi_0$. The background arrows show the direction field of the dynamical system. All trajectories converge rapidly to the slow-roll attractor and subsequently spiral toward the origin after inflation, demonstrating the global attractor behavior of the inflationary solution.} 
    \label{fig:phase_portrait}
\end{figure}

Moreover, Fig.~\ref{fig:phase_portrait} presents the full phase portrait of the model for the same parameter set as Fig.\ref{fig:phase_single}. In addition to the reference trajectory from Fig.~\ref{fig:phase_single}, showing by solid black line, we plot several other trajectories starting from the same initial field value $\phi_0$ but with different initial velocities $\dot\phi_0$. We also overlay the direction field of the dynamical system~\eqref{eq:phase1} and \eqref{eq:phase2}, shown by series of small arrows at each point in the $(\phi, \dot\phi)$ plane. We found three key features. First, all trajectories, regardless of their initial velocity, converge to a single curve in the $(\phi, \dot\phi)$ plane after a short transient. This single common (here the black line) is the slow-roll attractor, and the exponential decay of perturbations discussed analytically in Sec.~\ref{subsubsec:HJ} defines how rapidly they are going to converge. Second, the arrows of the direction field across the entire phase-space region all point toward the attractor, showing that the convergence to the slow-roll solution is a global property and not just restricted to nearby trajectories. Third, after the end of inflation, all trajectories merge into the same spiral motion toward the origin, indicating that the post-inflationary behavior is also independent of the initial conditions.

The above analysis confirms that the inflationary solution of the k-essence model with the Starobinsky potential and a power-law coupling is an attractor. Regardless of the initial field velocity, all trajectories converge rapidly to the same slow-roll solution, and the post-inflationary evolution is equally well-behaved. The coupling function $F(\phi)$ controls the rate at which trajectories converge to the attractor. This establishes that the model is robust and its dynamics do not depend on any special choice of initial conditions.

\section{Reheating}
\label{sec:reheating}
Once the inflation ends, the universe turns into an inert and supercooled state. To have a successful nucleosynthesis and produce the correct abundance of the light elements, the universe must undergo a transition from a supercooled state to a radiation-dominated phase after the end of inflation. A phase of reheating is required for this purpose; during this phase, the inflation field oscillates at the minimum of the potential and decays into other fields. When these fields interact with each other and reheat the universe, it leads to a universe filled with the relativistic degree of freedom~\cite{Albrecht:1982mp, Kofman:1994rk, Kofman:1997yn,
Bassett:2005xm, Rehagen:2015zma,Bhat:2020wdc}. The transition of the universe from supercooled to hot, thermal, and radiation-dominated states can occur by perturbative reheating or parametric resonance (preheating). In ~\cite{Cook:2015vqa}, the authors have developed a method that allows us to access the reheating epoch in a model-independent way without going to the complex dynamics of the reheating phase. The epoch of reheating can be studied with the help of three important parameters: $N_{re}$, the number of e-folds during the reheating phase), $T_{re}$ the thermalisation temperature, and the equation of states during reheating  $w_{re}$. Considering a constant $w_{\rm re}$ throughout reheating, the reheating e-fold number is given by~\cite{Cook:2015vqa, Martin:2014nya}
\begin{equation}
    N_{\rm re} = \frac{1}{3(1 + w_{\rm re})}
    \ln\!\left(\frac{90\,M_p^2 H_e^2}{\pi^2\,g_{\star\rm re}\,T_{\rm re}^4}\right),
    \label{eq:Nre}
\end{equation}
where $H_e$ is the Hubble parameter at the end of inflation and $g_{\star\rm re}$
denotes the number of relativistic degrees of freedom in the thermal bath. Adopting entropy conservation from the end of reheating to the present epoch and utilising Eq.~\eqref{eq:Nre}, we can write the reheating temperature~\cite{Cook:2015vqa}
\begin{eqnarray}
    T_{\rm re} = \left[ \left(\frac{11\,g_{\star s,\rm re}}{43}\right)^{1/3}
    \!\!\left(\frac{90\,M_p^2 H_e^2}{\pi^2\,g_{\star\rm re}}\right)^{\!\frac{1}{3(1+w_{\rm re})}}
     \,\frac{k_\star}{H_k\,T_0}\,e^{N_k}
    \right]^{\!\frac{3(1+w_{\rm re})}{1-3w_{\rm re}}},
    \label{eq:Tre}
\end{eqnarray}
where $H_k$ is the Hubble parameter at the time when the pivot scale $k_\star$ exits the horizon, $N_k$ is the number of e-folds of inflation, and $T_0 = 2.35\times10^{-4}$~eV is the current temperature of the universe. Using~\eqref{eq:Nre} and \eqref{eq:Tre}
we can denote the reheating in terms of inflationary observables.  Combining the above two equations, the reheating e-folds are obtained in terms of inflation e-folds as
\begin{eqnarray}\label{eqNre2}
    N_{\rm re} &=& \frac{4}{1-3w_{\rm re}} \; \left[ -N_k - \ln(\frac{k_\star}{T_0}) -  \frac{1}{3} \ln(\frac{11 g_{\star s}}{43})  - \frac{1}{4} \ln(\frac{90 M_p^2}{\pi^2 g_{\star re}}\frac{H_e^2}{H_\star^4}) \right]. 
\end{eqnarray}
The above equations show explicitly that the reheating temperature and e-folds depend on the inflationary e-folds $N_k$. Once $w_{\rm re}$ is fixed, both of these parameters are determined for given values of inflationary e-folds $N_k$ and the background quantities $H_k$ and $H_e$. Therefore, the value of $N_k$ used to predict the observable parameters $n_s$, $\alpha_s$, and $r$ in Sec.\ref{sec:results} also fix the reheating quantities.

\subsection*{Equation of state for the Starobinsky-like potential} 

After the end of inflation, the scalar field rolls down and reaches the minimum of its potential, i.e., $\phi_{\rm min} = 0$ depend, and begins to oscillate coherently around the minimum point. Practically, the equation of state during this oscillating condensate is not a fixed parameter, and in general, it depends on the shape of the potential, and for our case it may also depends on the coupling function. To determine the equation of state, we follow the standard procedure proposed in \cite{Turner:1983he}, which is based on the time-averaged energy density and pressure of the scalar field. From the action~\eqref{eq:action}, the energy density and pressure of the scalar field are
\begin{equation}
    \rho_\phi = \frac{F(\phi)}{2}\dot\phi^2 + V(\phi), \qquad
    p_\phi = \frac{F(\phi)}{2}\dot\phi^2 - V(\phi),
    \label{eq:rho_p}
\end{equation}
and the equation of state is defined as 
\begin{equation}\label{eq:w_inst}
    w_\phi \equiv \frac{p_\phi}{\rho_\phi}
    = \frac{F(\phi)\dot\phi^2 - 2V}{F(\phi)\dot\phi^2 + 2V}.
\end{equation}
The instantaneous equation of state $w_\phi = p_\phi / \rho_\phi$ oscillates time between $w_\phi = +1$ and $w_\phi = -1$ during each cycle. At the zero point of the field, all the energy is in kinetic form and $w_\phi = +1$, while at the turning point, all energy is in potential shape and $w_\phi = -1$. The effective reheating equation of state $w_{\rm re}$ is defined as the time-averaged value of $w_\phi$ over one oscillation cycle, i.e., $w_{\rm re} = \langle w_\phi \rangle$. For a canonical scalar field, where the potential has a minimum at $\phi = 0$, and the scalar field oscillate around this minimum, the time-averaged equation of state has a general expression given by~\cite{Turner:1983he}
\begin{equation}\label{eq:wre_integral}
    1 + w_{\rm re} = \frac{2\displaystyle\int_{0}^{\phi_m} 
    \left[1 - \frac{V(\phi)}{V(\phi_m)}\right]^{1/2} d\phi}
    {\displaystyle\int_{0}^{\phi_m} 
    \left[1 - \frac{V(\phi)}{V(\phi_m)}\right]^{-1/2} d\phi},
\end{equation}
where $\phi_m$ denotes the field amplitude at the turning point of the current oscillation cycle, so that it is the maximum value of the field during the cycle. It can be defined by the condition $\dot\phi = 0$, in which $\rho_\phi = V(\phi_m)$. Then, the integral depends on how much time the field spends near each extreme. 
For a potential that behaves as a power law near the minimum, i.e., $V(\phi) \propto \phi^p$, the integrals in Eq.~\eqref{eq:wre_integral} can be evaluated analytically, leading to the effective equation of state $w_{\rm re} = (p - 2)/(p + 2)$ ~\cite{Turner:1983he, Cook:2015vqa}. 
For the non-canonical model considered here, the derivation must be reconsidered due to the presence of the kinetic coupling. From Eq.\eqref{eq:w_inst}, it is found that $1 + \langle w_\phi \rangle = \langle F \dot\phi^2 / \rho_\phi \rangle$. Then, following the same procedure, and using the relation $\dot\phi^2 = 2(\rho_\phi - V)/F$, one can arrive at
\begin{equation}\label{eq:wre_noncanonical}
    1 + \langle w_\phi \rangle = 
    \frac{2\displaystyle\int_{\phi_m^-}^{\phi_m^+}
    \sqrt{F(\phi) \left( 1 - \dfrac{V(\phi)}{V(\phi_m)} \right)}
    \;d\phi}
    {\displaystyle\int_{\phi_m^-}^{\phi_m^+}
    \sqrt{F(\phi) \left( 1 - \dfrac{V(\phi)}{V(\phi_m)} \right)^{-1}}
    \;d\phi},
\end{equation}
in which $\phi_m^-$ and $\phi_m^+$ are the field values at the two turning points of the oscillation, where we have $V(\phi_\pm) = V(\phi_m) = \rho_\phi$. This is because the potential (Eq. \ref{eq:starobinsky}) is not symmetric around $\phi = 0$, so that, on the left side of the potential minimum ($\phi < 0$) compared to the left ($\phi > 0$), it can reach to the same hight for smaller field magnitude, i.e., $V(\phi_m^+) = V(\phi_M^-) = V_m$ where $|\phi_m^-| < |\phi_m^+|$. 

\begin{figure}
    \centering
    \includegraphics[width=0.5\linewidth]{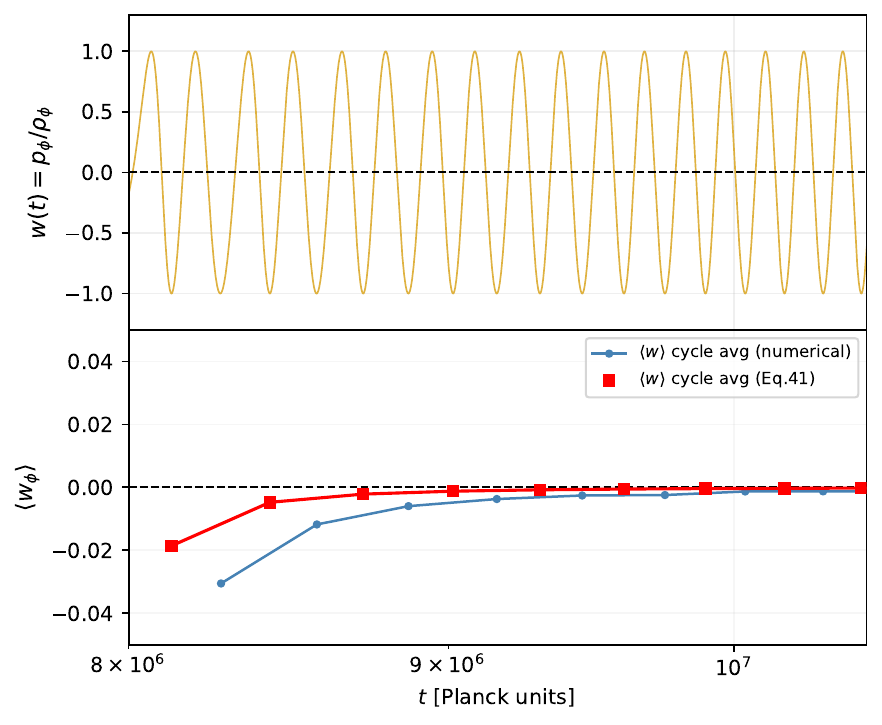}
    \caption{The figure shows the instantaneous (top) and cycle-averaged (bottom) equation of state of the scalar field during the reheating phase, for $A = 0.55$, $q = 2.0$. The effective equation of state converges to $w_{\rm re} \to 0$, consistent with the quadratic behavior of the Starobinsky potential near its minimum.}
    \label{fig:w_reheating}
\end{figure}

Note also that the $\sqrt{F(\phi)}$ factors do not cancel between the numerator and denominator. The coupling term affects the effective equation of state through the integration; however, it does not affect the extreme values of the instantaneous $w_\phi$ in which it still oscillates between $w=+1$ and $w=-1$; as shown in the top panel of Fig.\ref{fig:w_reheating}. The effective equation of state $w_{\rm re}$ in general depends on both the shape of the potential and the kinetic coupling function $F(\phi)$. To verify this, we evaluate Eq.\eqref{eq:wre_noncanonical} numerically with $F(\phi)$ retained inside the integrands. The turning points $\phi_m^\pm$ are determined numerically for each cycle by solving $V(\phi_m^\pm) = \rho_\phi$ relation. The results are shown by the bottom panel of Fig.\ref{fig:w_reheating}. The effective equation of state is obtained as $w_{\rm re} \simeq - 0.2$ for the first few cycles showing the effect of the kinetic coupling term at larger oscillation amplitude. Then, it converges to $w_{\rm re} \rightarrow  0$ as the amplitude of oscillation decays under friction and the contribution of the coupling term becomes small. Therefore, after a few cycles, the kinetic coupling term approaches $F(\phi) \to 1$ as the amplitude of the oscillation decreases and the effective equation of state is determined entirely by the shape of the potential near the minimum, and the standard relation $w_{\rm re} = (p - 2)/(p + 2)$ applies. Since the potential under consideration here (Eq. \ref{eq:starobinsky}) behaves as $V(\phi) \approx \phi^2$ near $\phi = 0$, the effective equation of state is estimated as $w_{\rm re} = 0$. 
\subsection*{Reheating temperature and constraints}

Once the $w_{\rm re} = 0$ is fixed, the reheating e-folds $N_{\rm re}$ and reheating temperature $T_{\rm re}$ are determined by the inflationary quantities $H_k$, $H_e$, $N_k$ and $n_s$ through Eqs.~\eqref{eq:Tre} and ~\eqref{eqNre2}, so that $N_{\rm re}$ and $T_{\rm re}$ are specified once is chosen, and are not free or independently adjustable quantities. In general, a longer inflationary period, corresponding to higher e-folds $N_k$, leads to a lower reheating temperature. Additionally, the faster reheating occurs the higher reheating temperature~\cite{Cook:2015vqa}.  

The reheating temperature is not directly measured by any observation, but it is constrained indirectly, since it should stand within a well-defined physically allowed range. The upper limit comes from the energy scale at the end of inflation. Determining the energy scale by the potential $V(\phi_e)$, the maximum reheating temperature corresponds to the case of instantaneous reheating, which gives $T_{\rm re}^{\rm max} \sim 10^{15}~{\rm GeV}$~\cite{Cook:2015vqa}. On the other hand, the lower bound on the reheating temperature arises from the BBN, which requires the universe to be in the radiation-dominant phase before the onset of the nucleosynthesis leading to $T_{\rm re} \gtrsim T_{\rm BBN} \simeq 4~{\rm MeV}$~\cite{Kawasaki:1999na, Kawasaki:2000en, Hasegawa:2019jsa}. These two limits define a wide viable range for the reheating temperature $T_{\rm re}$. Further studies show that an additional constraint on the reheating temperature can come from the contribution of the primordial gravitational waves (PGWs) to the effective number of relativistic species, $\Delta N_{\rm eff}$. It has been determined that the observational bound $\Delta N_{\rm eff} \leq 0.17$ at $95\%$ confidence~\cite{ACT:2025fju, ACT:2025tim} can translate into a lower limit on $T_{\rm re}$, which can be more stringent than the BBN bounds~\cite{Mohammadi:2025gbu, Haque:2025uri, Haque:2025uis}. However, the resulting new constraint is efficient when the reheating equation of state is stiffer than radiation, $w_{\rm re} > 1/3$~\cite{Boyle:2005se, Watanabe:2006qe, Saikawa:2018rcs, Figueroa:2019paj, Bernal:2019lpc}. Since in our model, the reheating equation of state is obtained to be $w_{\rm re} = 0$, this constraint does not apply, and the only relevant lower bound remain the BBN temperature.

\begin{figure}[h]
    \centering
    \includegraphics[width=0.5\linewidth]{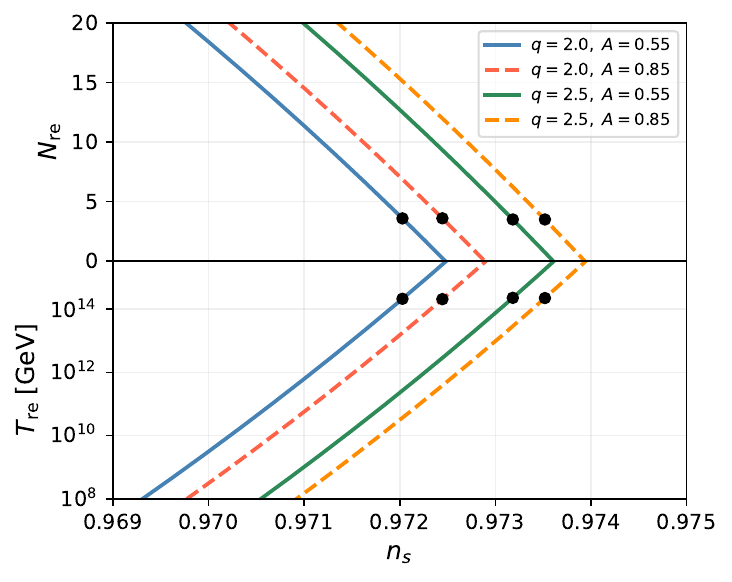}
    \caption{Reheating e-folds (top panel) and reheating temperature $T_{\rm re}$ as a function of the scalar spectral index $n_s$, obtained from Eq.~\eqref{eq:Tre} with $w_{\rm re} = 0$ and different values of the coupling parameters. The black dots, indicates the results corresponding to the inflationary e-folds $N_k = 55$. 
    }
    \label{fig:reheating}
\end{figure}
Fig.\ref{fig:reheating} presents the resulting reheating e-folds (top panel) and reheating temperature (bottom panel) versus the scalar spectral index for different sets of the coupling parameters, $n$ and $A$. The scalar spectral index $n_s$ is directly related to the inflationary e-folds $N_k$ through the slow-roll parameters, and we found that larger $n_s$ corresponds to larger $N_k$; as shown in Sec.\ref{sec:results}. The results determine that the physically acceptable inflationary e-folds must stay in the range $N_k \lesssim 58.2$, so that for larger $N_k$ results in negative reheating e-folds, which is physically unacceptable. The limiting case $N_{\rm re} = 0$ corresponds to instantaneous reheating, which gives the maximum reheating temperature as $T_{\rm re}^{\rm max} \simeq 10^{15}~{\rm GeV}$.
Although the inflationary observables are consistent with ACT DR6 for 
horizon crossing at $N_k = 55-60$ e-folds before the end of inflation, the study of reheating put an extra upper bound on the inflationary phase, so that $N_k \gtrsim 58.2$ leads to a negative $N_{\rm re}$ which excludes those solutions as physically inconsistent even when the predicted $n_s$, $\alpha_s$, and $r$ are consistent with the data. The black dots on the figure display the result corresponding to inflationary e-folds $N_k = 55$, for which
\begin{equation}\label{eq:Tre_range}
    N_{\rm re} \sim 3.5, \qquad T_{\rm re} \sim 10^{14}~{\rm GeV},
\end{equation}
well above the BBN bound, and confirms that the model produces a viable reheating phase for all parameter values consistent with the ACT DR6 $1\sigma$ region.

\section{Relic gravitational waves}
\label{sec:gw}
The stochastic PGW's background arises from the inflation that contains the entire history of the inflationary and post-inflationary epoch. During inflation, tensor perturbations in the metric provide a stochastic gravitational wave backdrop that exists as a relic signal to this day. In our model the scalar-field Lagrangian $\mathcal{L} = F(\phi)X - V(\phi)$ modifies only the scalar sector; the tensor perturbation equation remains the same,
\begin{equation}
    {h_\mathbf{k}^{\lambda}}'' + 2\mathcal{H}\,{h_\mathbf{k}^{\lambda}}'
    + k^2 h_\mathbf{k}^{\lambda} = 0,
    \label{eq:GW_eom}
\end{equation}
where $\mathcal{H} = a'/a$ and primes represent derivatives with respect to conformal time. The non-canonical coupling $F(\phi)$ does not play any active role in producing the PGWs. The amplitude of the PGWs is entirely dependent on the value of the Hubble parameter~\cite{Baumann:2009ds}.  
\begin{equation}
    \Delta_{T,\mathrm{prim}}^2(k)
    = \frac{2}{\pi^2}\left(\frac{H_{\mathrm{inf}}}{M_p}\right)^2,
    \label{eq:DeltaT}
\end{equation}re-enters,
summed over both polarisation states. After the horizon exit, the evolution of the PGWs freezes until the modes re-enter the post-inflationary era. Once the mode re-enters, the amplitude of the corresponding PGWs falls as  $a^{-1}$.
The fractional GW energy density today is~\cite{Kuroyanagi:2008ye}
\begin{equation}
    \Omega_{\mathrm{GW},0}(k)
    = \frac{1}{12}\frac{k^2}{a_0^2 H_0^2}
    \,\Delta_{T,\mathrm{prim}}^2(k)\cdot\mathcal{T}^2(k),
    \label{eq:OmegaGW_def}
\end{equation}
where  $a_\mathrm{hc}$ is the scale factor at horizon re-entry, and the transfer function is given, 
\begin{equation}
    \mathcal{T}^2(k) = \frac{1}{2}\left(\frac{a_\mathrm{hc}}{a_0}\right)^2,
    \label{eq:transfer}
\end{equation}
and the factor $1/2$ accounts for time-averaging over GW oscillations. The multiband structure in GWs arises due to the different ratios of $a_\mathrm{hc}/a_0$ that depend on different epochs.

Accepting the entropy conservation $g_{*s}T^3a^3 = \mathrm{const}$ along with the Friedmann equation $H^2 = \pi^2 g_*(T)\,T^4/(90\,M_p^2)$ for the \emph{radiation domination} phase, we can write 
\begin{equation}
    \frac{a_{\mathrm{hc},\mathrm{RD}}}{a_0}
    = \frac{a_0 H_0\sqrt{\Omega_{r0}}}{k}
    \left(\frac{g_*}{g_{*0}}\right)^{1/2}
    \!\!\left(\frac{g_{*s}}{g_{*s0}}\right)^{-2/3},
    \label{eq:ahc_RD}
\end{equation}
where we take into consideration the present values of  $g_{*0} = 3.36$ and $g_{*s0} = 3.91$. Also,
substituting Eq.~\eqref{eq:ahc_RD} into Eqs.~\eqref{eq:OmegaGW_def}
and~\eqref{eq:transfer}, the $k-$dependence cancels out, and we obtain
\emph{flat plateau}~\cite{Kuroyanagi:2008ye, Watanabe:2006qe, Ahmad:2019jbm}:
\begin{equation}
    \Omega_{\mathrm{GW},0}^{(\mathrm{RD})}
    = \frac{1}{6\pi^2}\,\Omega_{r0}
    \left(\frac{H_{\mathrm{inf}}}{M_p}\right)^2
    \left(\frac{g_*}{g_{*0}}\right)
    \!\!\left(\frac{g_{*s}}{g_{*s0}}\right)^{-4/3},
    \label{eq:OmegaRD}
\end{equation}
which is valid for $f_\mathrm{eq} < f < f_\mathrm{re}$. The two $g_*$ factors have different physical origins: the first derives from $H_\mathrm{hc}$ in the Friedmann equation, and the second from entropy conservation between the reheating period and now. For Standard Model values ($g_* = 106.75$), they combine to reduce the plateau by around $0.39$ compared to a naive calculation that ignores the fall in relativistic degrees of freedom.

Similarly, during \emph{matter domination}, $H^2 = H_0^2\Omega_{m0}(a_0/a)^3$, with a horizon-crossing scale factor that follows $a_\mathrm{hc} \propto k^{-2}$ (instead of $a^{-1}$ during radiation domination). During the matter-dominated phase, the energy budget is dominated by the non-relativistic matter; hence, it $g_*$ is absent. The spectra in this band may be described directly in terms of the radiation spectrum as
\begin{equation}\label{eq:OmegaMD}
    \Omega_{\mathrm{GW},0}^{(\mathrm{MD})}
    = \Omega_{\mathrm{GW},0}^{(\mathrm{RD})}
    \cdot\frac{\Omega_{m0}^2}{\Omega_{r0}}
    \left(\frac{g_*}{g_{*0}}\right)^{-1}
    \!\!\left(\frac{g_{*s}}{g_{*s0}}\right)^{4/3}
    \!\!\left(\frac{f_h}{f}\right)^2,
\end{equation}
$f_h = H_0/(2\pi)$ denotes the current Hubble frequency. The spectra in this band decay as $f^{-2}$, indicating the suppression of sub-horizon gravitational waves during matter domination.

In the \emph{reheating epoch}, modes with $f_\mathrm{re} < f < f_\mathrm{end}$
re-enter while the inflaton is oscillating around the minima of its potential. The PGW's spectrum in for a general equation of state $w_\mathrm{re}$, can be written as~\cite{Boyle:2005se, Watanabe:2006qe,Ahmad:2019jbm}
\begin{equation}
    \Omega_{\mathrm{GW},0}^{(\mathrm{RE})}
    = \Omega_{\mathrm{GW},0}^{(\mathrm{RD})}
    \left(\frac{f_\mathrm{re}}{f}\right)^{\!\frac{2(1-3w_\mathrm{re})}{1+3w_\mathrm{re}}}.
    \label{eq:OmegaRE}
\end{equation}
The spectral slope in the reheating band is exclusively determined by $w_\mathrm{re}$. In quintessential inflation, a stiff equation of state $w_\mathrm{re} > 1/3$, such kinetic dominance ($w_\mathrm{re} = 1$), results in a blue-tilted spectrum. In contrast, in the model under discussion, in our model (Eq. \ref{eq:starobinsky}) with $w_\mathrm{re} = 0$, the exponent term becomes $2$, and the spectrum is suppressed as
\begin{equation}
    \Omega_{\mathrm{GW},0}^{(\mathrm{RE})}
    = \Omega_{\mathrm{GW},0}^{(\mathrm{RD})}
    \left(\frac{f_\mathrm{re}}{f}\right)^2,
    \qquad (f_\mathrm{re} < f \leq f_\mathrm{end}),
    \label{eq:OmegaRE_w0}
\end{equation}
a direct consequence of the energy density redshifting as $a^{-3}$ during the matter-like reheating phase, which is slower than $a^{-4}$ scaling for the radiation phase. This is the central qualitative difference between
our scenario and quintessential inflation models, where the kinetic regime
generates a pronounced blue peak in the GW spectrum \cite{Ahmad:2019jbm}.

The full spectrum is therefore~\cite{Boyle:2005se, Watanabe:2006qe,Ahmad:2019jbm}
\begin{widetext}
\begin{equation}
\Omega_{\mathrm{GW},0}(f)\,h^2 = \Omega_{\mathrm{GW},0}^{(\mathrm{RD})}\,h^2 \times
\begin{cases}
\dfrac{\Omega_{m0}^2}{\Omega_{r0}}
\!\left(\dfrac{g_*}{g_{*0}}\right)^{\!-1}
\!\!\left(\dfrac{g_{*s}}{g_{*s0}}\right)^{\!4/3}
\!\!\left(\dfrac{f_h}{f}\right)^2
& f_h < f \leq f_{\rm eq}, \\[10pt]
1
& f_{\rm eq} < f \leq f_{\rm re}, \\[6pt]
\!\left(\dfrac{f_{\rm re}}{f}\right)^2
& f_{\rm re} < f \leq f_{\rm end}, \\[6pt]
0 & f > f_{\rm end},
\end{cases}
\label{eq:OmegaFull}
\end{equation}
\end{widetext}
where the transition frequencies are~\cite{Kuroyanagi:2008ye}
\begin{equation}\label{eq:feq}
    f_\mathrm{eq} = \frac{\sqrt{2}\,H_0}{2\pi}
    \frac{\Omega_{m0}}{\sqrt{\Omega_{r0}}}
    \simeq 1.6\times10^{-17}\ \mathrm{Hz},
\end{equation}
\begin{equation}\label{eq:fre}
    f_\mathrm{re} = \frac{(43/11)^{1/3}\,g_*^{1/6}\,T_0\,T_\mathrm{re}}
    {6\sqrt{10}\,M_p}
    \simeq 1.6\times10^3
    \!\left(\frac{T_\mathrm{re}}{10^{10}\,\mathrm{GeV}}\right)\mathrm{Hz},
\end{equation}
\begin{equation}\label{eq:fend}
    f_\mathrm{end} \simeq 8.9\times10^8
    \!\left(\frac{H_\mathrm{end}}{10^{14}\,\mathrm{GeV}}\right)^{\!1/2}
    \mathrm{Hz}.
\end{equation}

\begin{figure}[t]
    \centering
    \includegraphics[width=0.5\linewidth]{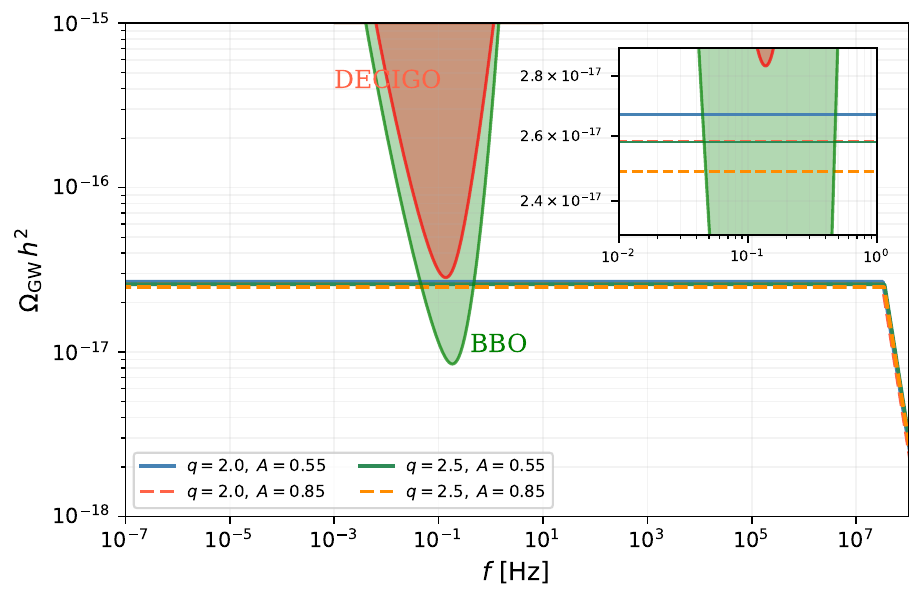}
    \caption{Energy spectrum of primordial gravitational waves     $\Omega_{\mathrm{GW}}\,h^2$ as a function of frequency is displayed for     $N_k = 55$ and different values of the coupling parameters $n$ and $A$. The sensitivity curves of DECIGO and BBO are shown for comparison.     Since $w_\mathrm{re} = 0$ for the Starobinsky potential, the spectrum falls as $f^{-2}$ in the reheating frequency window ($f_\mathrm{re} < f < f_\mathrm{end}$) rather than being enhanced, in contrast to quintessential inflation scenarios where kinetic domination produces a blue-tilted peak \cite{Ahmad:2019jbm}.}
    \label{fig:GW}
\end{figure}
We obtained a tensor-to-scalar ratio $r \lesssim 0.006$, for a wide range of parameter space, which is consistent with the latest ACT observation. This results in a plateau in the amplitude of the GWs $\Omega_\mathrm{GW}h^2 \sim 10^{-17}$.  Such a low value of the amplitude is well below the current detector range. The GW energy spectrum~\eqref{eq:OmegaFull} is plotted in Fig.~\ref{fig:GW} for two distinct values of the number of e-folds during inflation, $N_k = 55$ and $N_k = 60$. The two curves have the same flat plateau but differ in the position of $f_\mathrm{re}$, which changes with the reheating temperature. A higher $N_k$ results in a lower $T_\mathrm{re}$, which leads to a lower $f_\mathrm{re}$, extending the suppressed reheating band to lower frequencies. The results we obtained here demonstrate that there is no enhancement in the GWs' amplitude for $w_\mathrm{re} = 0$. It is evident from figure~\ref {fig:GW} that only BBO can detect such signals in the near future.

\section{Conclusion}
\label{sec:conclusion}

In this work, we studied k-essence inflation with the Lagrangian $\mathcal{L} = F(\phi)X - V(\phi)$, where the potential is taken as a Starobinsky-like form and the coupling function is assumed to be described by a power-law function, $F(\phi) = 1 + A\phi^n$. This model is theoretically well-motivated so that it emerges naturally as a special case of the G-inflation framework, which is itself a sub-sector of the Horndeski scalar-tensor theory. Due to the linear dependency of the Lagrangian to the kinetic term $X$, the sound speed of curvature perturbations is exactly unity, the same as the canonical model. The observational motivation for this study is the recent ACT DR6 data release, which disfavours the Starobinsky potential in the standard canonical framework. We investigated whether this simple and theoretically well-motivated modification of the kinetic sector, with no change to the gravitational theory, is sufficient to bring a Starobinsky-like potential into agreement with the ACT DR6 constraints.

The background equations were solved numerically, and the inflationary observables $n_s$, $\alpha_s$, and $r$ were evaluated at $N = 55$ and $N = 60$ e-folds before the end of inflation. The effects of the coupling parameters $A$ and $n$ on the observable parameters were studied, and it was found that increasing the coupling parameter raises the total number of inflationary e-folds, forcing the scalar field at horizon crossing to a smaller value. Increasing the coupling parameter $A$ or $n$ leads to a higher effective coupling strength $F(\phi_\star)$ and the ratio $V_{,\phi}/V$ at horizon crossing. While a larger coupling tends to suppress the slow-roll parameter $\epsilon_V$, the simultaneous increase in $V_{,\phi}/V$ has the opposite tendency so that it desires to rise $\epsilon_V$. The results showed that the ratio $V_{,\phi}/V$ has the dominant effect. Including both contributions together, the net result is that $\epsilon_V$ increases with the coupling parameters, leading to a higher $r$. At the same time, the combined effect on the remaining terms in $n_s$ results in a higher scalar spectral index as well. Despite this non-trivial behaviour, the resulting $n_s$, $\alpha_s$, and $r$ fall within the $1\sigma$ region of ACT DR6 across the full parameter space explored. This confirms that the potential mentioned in Eq. \ref{eq:starobinsky} in the k-essence framework can be counted as a viable candidate for inflation.

The attractor behaviour was verified both analytically through the Hamilton-Jacobi formalism and numerically by performing a phase-space analysis. In the analytical formalism, it is understood that due to the sign of the Hubble parameter $H$, its derivative $H_{_,\phi}$, and also because the coupling function is positive during the whole inflationary phase, the perturbations decay exponentially as the scalar field rolls down its potential and we approach the end of inflation. The presence of the coupling function $F(\phi)$ in the exponential term shows that it controls the rate of convergence. This was confirmed by doing a numerical exploration over the phase space we follow the trajectories starting from different initial field velocities. It was found that all these trajectories converge to the same slow-roll solution. After inflation, the trajectories merge and form a spiral motion around the origin. This result indicates that the model is robust and its predictions are not sensitive to the choice of initial conditions.

Then, the reheating phase was considered, where the reheating temperature and e-folds were discussed. After inflation, the inflaton oscillates around the minimum of the potential, i.e., $\phi = 0$. Due to the nature of the Starobinsky-like potential, it was discussed that the effective equation of state of reheating is fixed to $w_{\rm re} = 0$. Including this, the reheating temperature is found to be of the order of $T_{\rm re} \sim 10^{14}~{\rm GeV}$, well above the BBN bound, and the reheating e-folds are found to be around $N_{\rm re} \sim 5$. The relic GW spectrum was also computed, showing a suppression as $f^{-2}$ in the reheating frequency window and a plateau amplitude of $\Omega_{\rm GW}h^2 \sim 10^{-17}$, which can lie within the sensitivity bound of future detectors, i.e., BBO.

These results found in this work demonstrate that the k-essence inflation with a Starobinsky-like potential is a theoretically well-motivated and observationally consistent scenario. A simple power-law modification to the kinetic part, which naturally put the model as a subclass of G-inflation, is sufficient to bring a Starobinsky-like potential into agreement with ACT DR6.

\acknowledgments
MS is supported by the Science and Engineering Research Board (SERB), DST, Government of India, under the grant agreement number CRG/2022/004120 (Core Research Grant). MS is also partially supported by the Ministry of Education and Science of the Republic of Kazakhstan, Grant No. 0118RK00935.


\bibliography{kinflation_final}

@article{ACT:2025fju,
    author = "Louis, Thibaut and others",
    collaboration = "ACT",
    title = "{The Atacama Cosmology Telescope: DR6 Power Spectra, Likelihoods and $\Lambda$CDM Parameters}",
    eprint = "2503.14452",
    archivePrefix = "arXiv",
    primaryClass = "astro-ph.CO",
    journal = "{arXiv preprint}", 
    reportNumber = "FERMILAB-PUB-25-0071-PPD",
    month = "3",
    year = "2025"
}

@article{ACT:2025tim,
    author = "Calabrese, Erminia and others",
    collaboration = "ACT",
    title = "{The Atacama Cosmology Telescope: DR6 Constraints on Extended Cosmological Models}",
    eprint = "2503.14454",
    archivePrefix = "arXiv",
    primaryClass = "astro-ph.CO",
    journal = "{arXiv preprint}",
    volume = "{N/A}",  
    pages = "{N/A}",   
    reportNumber = "FERMILAB-PUB-25-0157-PPD",
    month = "3",
    year = "2025"
}

@article{Ahmad:2019jbm,
    author = "Ahmad, Safia and De Felice, Antonio and Jaman, Nur and Kuroyanagi, Sachiko and Sami, M.",
    title = "{Baryogenesis in the paradigm of quintessential inflation}",
    eprint = "1908.03742",
    archivePrefix = "arXiv",
    primaryClass = "gr-qc",
    reportNumber = "YITP-19-133",
    doi = "10.1103/PhysRevD.100.103525",
    journal = "Phys. Rev. D",
    volume = "100",
    number = "10",
    pages = "103525",
    year = "2019"
}

@article{Albrecht:1982wi,
    author = "Albrecht, Andreas and Steinhardt, Paul J.",
    editor = "Fang, Li-Zhi and Ruffini, R.",
    title = "{Cosmology for Grand Unified Theories with Radiatively Induced Symmetry Breaking}",
    reportNumber = "UPR-0185T",
    doi = "10.1103/PhysRevLett.48.1220",
    journal = "Phys. Rev. Lett.",
    volume = "48",
    pages = "1220--1223",
    year = "1982"
}

@article{Aoki:2025wld,
    author = "Aoki, Shuntaro and Otsuka, Hajime and Yanagita, Ryota",
    title = "{Higgs-modular inflation}",
    eprint = "2504.01622",
    archivePrefix = "arXiv",
    primaryClass = "hep-ph",
    reportNumber = "RIKEN-iTHEMS-Report-25, KYUSHU-HET-317",
    doi = "10.1103/v4z9-676d",
    journal = "Phys. Rev. D",
    volume = "112",
    number = "4",
    pages = "043505",
    year = "2025"
}

@article{Armendariz-Picon:1999hyi,
    author = "Armendariz-Picon, C. and Damour, T. and Mukhanov, Viatcheslav F.",
    title = "{k - inflation}",
    eprint = "hep-th/9904075",
    archivePrefix = "arXiv",
    doi = "10.1016/S0370-2693(99)00603-6",
    journal = "Phys. Lett. B",
    volume = "458",
    pages = "209--218",
    year = "1999"
}

@article{Armendariz-Picon:2000ulo,
    author = "Armendariz-Picon, C. and Mukhanov, Viatcheslav F. and Steinhardt, Paul J.",
    title = "{Essentials of k essence}",
    eprint = "astro-ph/0006373",
    archivePrefix = "arXiv",
    doi = "10.1103/PhysRevD.63.103510",
    journal = "Phys. Rev. D",
    volume = "63",
    pages = "103510",
    year = "2001"
}

@article{Vagnozzi:2022qmc,
    author = "Vagnozzi, Sunny and Loeb, Abraham",
    title = "{The Challenge of Ruling Out Inflation via the Primordial Graviton Background}",
    eprint = "2208.14088",
    archivePrefix = "arXiv",
    primaryClass = "astro-ph.CO",
    doi = "10.3847/2041-8213/ac9b0e",
    journal = "Astrophys. J. Lett.",
    volume = "939",
    number = "2",
    pages = "L22",
    year = "2022"
}

@article{BICEP:2021xfz,
    author = "Ade, P. A. R. and others",
    collaboration = "BICEP, Keck",
    title = "{Improved Constraints on Primordial Gravitational Waves using Planck, WMAP, and BICEP/Keck Observations through the 2018 Observing Season}",
    eprint = "2110.00483",
    archivePrefix = "arXiv",
    primaryClass = "astro-ph.CO",
    doi = "10.1103/PhysRevLett.127.151301",
    journal = "Phys. Rev. Lett.",
    volume = "127",
    number = "15",
    pages = "151301",
    year = "2021"
}

@article{Barenboim:2007ii,
      author         = "Barenboim, Gabriela and Kinney, William H.",
      title          = "{Slow roll in simple non-canonical inflation}",
      journal        = "JCAP",
      volume         = "0703",
      year           = "2007",
      pages          = "014",
      doi            = "10.1088/1475-7516/2007/03/014",
      eprint         = "astro-ph/0701343",
      archivePrefix  = "arXiv",
      primaryClass   = "astro-ph",
      SLACcitation   = "%%CITATION = ASTRO-PH/0701343;%%",
    number         = {},
}

@inproceedings{Baumann:2009ds,
    author = "Baumann, Daniel",
    title = "{Inflation}",
    booktitle = "{Theoretical Advanced Study Institute in Elementary Particle Physics}: {Physics of the Large and the Small}",
    eprint = "0907.5424",
    archivePrefix = "arXiv",
    primaryClass = "hep-th",
    reportNumber = "TASI-2009",
    doi = "10.1142/9789814327183_0010",
    pages = "523--686",
    year = "2011"
}

@article{Bernal:2019lpc,
    author = "Bernal, Nicol{\'a}s and Hajkarim, Fazlollah",
    title = "{Primordial Gravitational Waves in Nonstandard Cosmologies}",
    eprint = "1905.10410",
    archivePrefix = "arXiv",
    primaryClass = "astro-ph.CO",
    doi = "10.1103/PhysRevD.100.063502",
    journal = "Phys. Rev. D",
    volume = "100",
    number = "6",
    pages = "063502",
    year = "2019"
}

@article{Boyle:2005se,
    author = "Boyle, Latham A. and Steinhardt, Paul J.",
    title = "{Probing the early universe with inflationary gravitational waves}",
    eprint = "astro-ph/0512014",
    archivePrefix = "arXiv",
    doi = "10.1103/PhysRevD.77.063504",
    journal = "Phys. Rev. D",
    volume = "77",
    pages = "063504",
    year = "2008"
}

@unpublished{Brahma:2025dio,
    author = "Brahma, Suddhasattwa and Calder\'on-Figueroa, Jaime",
    title = "{Is the CMB revealing signs of pre-inflationary physics?}",
    eprint = "2504.02746",
    archivePrefix = "arXiv",
    primaryClass = "astro-ph.CO",
    month = "4",
    year = "2025"
}

@article{Chen:2006nt,
    author = "Chen, Xingang and Huang, Min-xin and Kachru, Shamit and Shiu, Gary",
    title = "{Observational signatures and non-Gaussianities of general single field inflation}",
    eprint = "hep-th/0605045",
    archivePrefix = "arXiv",
    reportNumber = "SLAC-PUB-11840, MAD-TH-06-3, UFIFT-HEP-06-9, SU-ITP-06-12, CU-TP-1147",
    doi = "10.1088/1475-7516/2007/01/002",
    journal = "JCAP",
    volume = "01",
    pages = "002",
    year = "2007"
}

@article{Cook:2015vqa,
    author = "Cook, Jessica L. and Dimastrogiovanni, Emanuela and Easson, Damien A. and Krauss, Lawrence M.",
    title = "{Reheating predictions in single field inflation}",
    eprint = "1502.04673",
    archivePrefix = "arXiv",
    primaryClass = "astro-ph.CO",
    doi = "10.1088/1475-7516/2015/04/047",
    journal = "JCAP",
    volume = "04",
    pages = "047",
    year = "2015",
    number         = {},
}

@article{Dioguardi:2025mpp,
    author = "Dioguardi, Christian and Karam, Alexandros",
    title = "{Palatini linear attractors are back in action}",
    eprint = "2504.12937",
    archivePrefix = "arXiv",
    primaryClass = "gr-qc",
    doi = "10.1103/23b3-9d7q",
    journal = "Phys. Rev. D",
    volume = "111",
    number = "12",
    pages = "123521",
    year = "2025"
}

@article{Dioguardi:2025vci,
    author = "Dioguardi, Christian and Iovino, Antonio J. and Racioppi, Antonio",
    title = "{Fractional attractors in light of the latest ACT observations}",
    eprint = "2504.02809",
    archivePrefix = "arXiv",
    primaryClass = "gr-qc",
    doi = "10.1016/j.physletb.2025.139664",
    journal = "Phys. Lett. B",
    volume = "868",
    pages = "139664",
    year = "2025"
}

@article{Drees:2025ngb,
    author = "Drees, Manuel and Xu, Yong",
    title = "{Refined predictions for Starobinsky inflation and post-inflationary constraints in light of ACT}",
    eprint = "2504.20757",
    archivePrefix = "arXiv",
    primaryClass = "astro-ph.CO",
    reportNumber = "MITP-25-033",
    doi = "10.1016/j.physletb.2025.139612",
    journal = "Phys. Lett. B",
    volume = "867",
    pages = "139612",
    year = "2025"
}

@article{Ellis:2025zrf,
    author = "Ellis, John and Garcia, Marcos A. G. and Olive, Keith A. and Verner, Sarunas",
    title = "{Constraints on attractor models of inflation and reheating from Planck, BICEP/Keck, ACT DR6, and SPT-3G data}",
    eprint = "2510.18656",
    archivePrefix = "arXiv",
    primaryClass = "hep-ph",
    reportNumber = "UMN-TH-4512/25, FTPI-MINN-25/14, KCL-PH-TH/2025-42, CERN-TH-2025-199",
    doi = "10.1103/d35r-7bn8",
    journal = "Phys. Rev. D",
    volume = "113",
    number = "6",
    pages = "063571",
    year = "2026"
}

@article{Figueroa:2019paj,
    author = "Figueroa, Daniel G. and Tanin, Erwin H.",
    title = "{Ability of LIGO and LISA to probe the equation of state of the early Universe}",
    eprint = "1905.11960",
    archivePrefix = "arXiv",
    primaryClass = "astro-ph.CO",
    doi = "10.1088/1475-7516/2019/08/011",
    journal = "JCAP",
    volume = "08",
    pages = "011",
    year = "2019",
    number         = {},
}

@article{Franche:2010yj,
      author         = "Franche, Paul and Gwyn, Rhiannon and Underwood, Bret and
                        Wissanji, Alisha",
      title          = "{Initial Conditions for Non-Canonical Inflation}",
      journal        = "Phys. Rev.",
      volume         = "D82",
      year           = "2010",
      pages          = "063528",
      doi            = "10.1103/PhysRevD.82.063528",
      eprint         = "arXiv:1002.2639",
      archivePrefix  = "arXiv",
      primaryClass   = "hep-th",
      SLACcitation   = "%%CITATION = ARXIV:1002.2639;%%"
}

@article{Gao:2025onc,
    author = "Gao, Qing and Gong, Yungui and Yi, Zhu and Zhang, Fengge",
    title = "{Nonminimal coupling in light of ACT data}",
    eprint = "2504.15218",
    archivePrefix = "arXiv",
    primaryClass = "astro-ph.CO",
    doi = "10.1016/j.dark.2025.102106",
    journal = "Phys. Dark Univ.",
    volume = "50",
    pages = "102106",
    year = "2025"
}

@article{Garriga:1999vw,
      author         = "Garriga, Jaume and Mukhanov, Viatcheslav F.",
      title          = "{Perturbations in k-inflation}",
      journal        = "Phys. Lett.",
      volume         = "B458",
      year           = "1999",
      pages          = "219-225",
      doi            = "10.1016/S0370-2693(99)00602-4",
      eprint         = "hep-th/9904176",
      archivePrefix  = "arXiv",
      primaryClass   = "hep-th",
      reportNumber   = "UAB-FT-466",
      SLACcitation   = "%%CITATION = HEP-TH/9904176;%%"
}

@unpublished{Gialamas:2025kef,
    author = "Gialamas, Ioannis D. and Karam, Alexandros and Racioppi, Antonio and Raidal, Martti",
    title = "{Has ACT measured radiative corrections to the tree-level Higgs-like inflation?}",
    eprint = "2504.06002",
    archivePrefix = "arXiv",
    primaryClass = "astro-ph.CO",
    month = "4",
    year = "2025"
}

@article{Gialamas:2025ofz,
    author = "Gialamas, Ioannis D. and Katsoulas, Theodoros and Tamvakis, Kyriakos",
    title = "{Keeping the relation between the Starobinsky model and no-scale supergravity ACTive}",
    eprint = "2505.03608",
    archivePrefix = "arXiv",
    primaryClass = "gr-qc",
    doi = "10.1088/1475-7516/2025/09/060",
    journal = "JCAP",
    volume = "09",
    pages = "060",
    year = "2025",
    number         = {},
}

@article{Guo:2003nz,
    author = "Guo, Zong-Kuan and Zhang, Hong-Sheng and Zhang, Yuan-Zhong",
    title = "{Inflationary attractor in brane world scenario}",
    eprint = "hep-ph/0309163",
    archivePrefix = "arXiv",
    doi = "10.1103/PhysRevD.69.063502",
    journal = "Phys. Rev. D",
    volume = "69",
    pages = "063502",
    year = "2004"
}

@article{Guth:1980zm,
      author         = "Guth, Alan H.",
      title          = "{The Inflationary Universe: A Possible Solution to the
                        Horizon and Flatness Problems}",
      journal        = "Phys. Rev.",
      volume         = "D23",
      year           = "1981",
      pages          = "347-356",
      doi            = "10.1103/PhysRevD.23.347",
      note           = "[Adv. Ser. Astrophys. Cosmol.3,139(1987)]",
      reportNumber   = "SLAC-PUB-2576",
      SLACcitation   = "%%CITATION = PHRVA,D23,347;%%"
}

@article{Guth:1982ec,
    author = "Guth, Alan H. and Pi, S. Y.",
    title = "{Fluctuations in the New Inflationary Universe}",
    doi = "10.1103/PhysRevLett.49.1110",
    journal = "Phys. Rev. Lett.",
    volume = "49",
    pages = "1110--1113",
    year = "1982"
}

@unpublished{Haque:2025uis,
    author = "Haque, Md Riajul and Pal, Sourav and Paul, Debarun",
    title = "{Improved Predictions on Higgs-Starobinsky Inflation and Reheating with ACT DR6 and Primordial Gravitational Waves}",
    eprint = "2505.04615",
    archivePrefix = "arXiv",
    primaryClass = "astro-ph.CO",
    month = "5",
    year = "2025"
}

@unpublished{Haque:2025uri,
    author = "Haque, Md Riajul and Pal, Sourav and Paul, Debarun",
    title = "{ACT DR6 Insights on the Inflationary Attractor models and Reheating}",
    eprint = "2505.01517",
    archivePrefix = "arXiv",
    primaryClass = "astro-ph.CO",
    month = "5",
    year = "2025"
}

@article{Hasegawa:2019jsa,
    author = "Hasegawa, Takuya and Hiroshima, Nagisa and Kohri, Kazunori and Hansen, Rasmus S. L. and Tram, Thomas and Hannestad, Steen",
    title = "{MeV-scale reheating temperature and thermalization of oscillating neutrinos by radiative and hadronic decays of massive particles}",
    eprint = "1908.10189",
    archivePrefix = "arXiv",
    primaryClass = "hep-ph",
    reportNumber = "KEK-TH-2149, KEK-Cosmo-242, RIKEN-iTHEMS-Report-19, IPMU19-0120",
    doi = "10.1088/1475-7516/2019/12/012",
    journal = "JCAP",
    volume = "12",
    pages = "012",
    year = "2019"
}

@article{Horndeski:1974wa,
    author = "Horndeski, Gregory Walter",
    title = "{Second-order scalar-tensor field equations in a four-dimensional space}",
    doi = "10.1007/BF01807638",
    journal = "Int. J. Theor. Phys.",
    volume = "10",
    pages = "363--384",
    year = "1974"
}

@article{Huang:2013hsb,
    author = "Huang, Qing-Guo",
    title = "{A polynomial f(R) inflation model}",
    eprint = "1309.3514",
    archivePrefix = "arXiv",
    primaryClass = "hep-th",
    doi = "10.1088/1475-7516/2014/02/035",
    journal = "JCAP",
    volume = "02",
    pages = "035",
    year = "2014",
    number         = {},
}

@article{Kallosh:2013yoa,
    author = "Kallosh, Renata and Linde, Andrei and Roest, Diederik",
    title = "{\em Superconformal Inflationary $\alpha$-Attractors}",
    eprint = "1311.0472",
    archivePrefix = "arXiv",
    primaryClass = "hep-th",
    doi = "10.1007/JHEP11(2013)198",
    journal = "JHEP",
    volume = "11",
    pages = "198",
    year = "2013",
    number         = {},
}

@article{Kallosh:2025rni,
    author = "Kallosh, Renata and Linde, Andrei and Roest, Diederik",
    title = "{Atacama Cosmology Telescope, South Pole Telescope, and Chaotic Inflation}",
    eprint = "2503.21030",
    archivePrefix = "arXiv",
    primaryClass = "hep-th",
    doi = "10.1103/d6gn-78hn",
    journal = "Phys. Rev. Lett.",
    volume = "135",
    number = "16",
    pages = "161001",
    year = "2025"
}

@article{Kamada:2010qe,
    author = "Kamada, Kohei and Kobayashi, Tsutomu and Yamaguchi, Masahide and Yokoyama, Jun'ichi",
    title = "{Higgs G-inflation}",
    eprint = "1012.4238",
    archivePrefix = "arXiv",
    primaryClass = "astro-ph.CO",
    reportNumber = "RESCEU-28-10",
    doi = "10.1103/PhysRevD.83.083515",
    journal = "Phys. Rev. D",
    volume = "83",
    pages = "083515",
    year = "2011"
}

@article{Kawasaki:1999na,
    author = "Kawasaki, M. and Kohri, Kazunori and Sugiyama, Naoshi",
    title = "{Cosmological constraints on late time entropy production}",
    eprint = "astro-ph/9811437",
    archivePrefix = "arXiv",
    reportNumber = "RESCEU-7-99, KUNS-1546",
    doi = "10.1103/PhysRevLett.82.4168",
    journal = "Phys. Rev. Lett.",
    volume = "82",
    pages = "4168",
    year = "1999"
}

@article{Kawasaki:2000en,
    author = "Kawasaki, M. and Kohri, Kazunori and Sugiyama, Naoshi",
    title = "{MeV scale reheating temperature and thermalization of neutrino background}",
    eprint = "astro-ph/0002127",
    archivePrefix = "arXiv",
    doi = "10.1103/PhysRevD.62.023506",
    journal = "Phys. Rev. D",
    volume = "62",
    pages = "023506",
    year = "2000"
}

@unpublished{Kim:2025dyi,
    author = "Kim, Jinsu and Wang, Xinpeng and Zhang, Ying-li and Ren, Zhongzhou",
    title = "{Enhancement of primordial curvature perturbations in $R^3$-corrected Starobinsky-Higgs inflation}",
    eprint = "2504.12035",
    archivePrefix = "arXiv",
    primaryClass = "astro-ph.CO",
    month = "4",
    year = "2025"
}

@article{Kinney:1997ne,
      author         = "Kinney, William H.",
      title          = "{A Hamilton-Jacobi approach to nonslow roll inflation}",
      journal        = "Phys. Rev.",
      volume         = "D56",
      year           = "1997",
      pages          = "2002-2009",
      doi            = "10.1103/PhysRevD.56.2002",
      eprint         = "hep-ph/9702427",
      archivePrefix  = "arXiv",
      primaryClass   = "hep-ph",
      reportNumber   = "FERMILAB-PUB-97-042-A",
      SLACcitation   = "%%CITATION = HEP-PH/9702427;%%"
}

@article{Kobayashi:2010cm,
    author = "Kobayashi, Tsutomu and Yamaguchi, Masahide and Yokoyama, Jun'ichi",
    title = "{G-inflation: Inflation driven by the Galileon field}",
    eprint = "1008.0603",
    archivePrefix = "arXiv",
    primaryClass = "hep-th",
    reportNumber = "RESCEU-18-10",
    doi = "10.1103/PhysRevLett.105.231302",
    journal = "Phys. Rev. Lett.",
    volume = "105",
    pages = "231302",
    year = "2010"
}

@article{Kuroyanagi:2008ye,
    author = "Kuroyanagi, Sachiko and Chiba, Takeshi and Sugiyama, Naoshi",
    title = "{Precision calculations of the gravitational wave background spectrum from inflation}",
    eprint = "0804.3249",
    archivePrefix = "arXiv",
    primaryClass = "astro-ph",
    doi = "10.1103/PhysRevD.79.103501",
    journal = "Phys. Rev. D",
    volume = "79",
    pages = "103501",
    year = "2009"
}

@article{Liddle:1994dx,
    author = "Liddle, Andrew R. and Parsons, Paul and Barrow, John D.",
    title = "{Formalizing the slow roll approximation in inflation}",
    eprint = "astro-ph/9408015",
    archivePrefix = "arXiv",
    reportNumber = "SUSSEX-AST-94-8-1",
    doi = "10.1103/PhysRevD.50.7222",
    journal = "Phys. Rev. D",
    volume = "50",
    pages = "7222--7232",
    year = "1994"
}

@article{Lin:2020goi,
    author = "Lin, Jiong and Gao, Qing and Gong, Yungui and Lu, Yizhou and Zhang, Chao and Zhang, Fengge",
    title = "{Primordial black holes and secondary gravitational waves from $k$ and $G$ inflation}",
    eprint = "2001.05909",
    archivePrefix = "arXiv",
    primaryClass = "gr-qc",
    doi = "10.1103/PhysRevD.101.103515",
    journal = "Phys. Rev. D",
    volume = "101",
    number = "10",
    pages = "103515",
    year = "2020"
}

@article{Linde:1981mu,
    author = "Linde, Andrei D.",
    editor = "Fang, Li-Zhi and Ruffini, R.",
    title = "{A New Inflationary Universe Scenario: A Possible Solution of the Horizon, Flatness, Homogeneity, Isotropy and Primordial Monopole Problems}",
    reportNumber = "LEBEDEV-81-229",
    doi = "10.1016/0370-2693(82)91219-9",
    journal = "Phys. Lett. B",
    volume = "108",
    pages = "389--393",
    year = "1982"
}

@article{Linde:1983gd,
    author = "Linde, Andrei D.",
    title = "{Chaotic Inflation}",
    doi = "10.1016/0370-2693(83)90837-7",
    journal = "Phys. Lett. B",
    volume = "129",
    pages = "177--181",
    year = "1983"
}

@article{Liu:2025qca,
    author = "Liu, Lang and Yi, Zhu and Gong, Yungui",
    title = "{Reconciling Nonminimally Coupled Higgs Inflation with ACT DR6 Observations through Reheating}",
    eprint = "2505.02407",
    archivePrefix = "arXiv",
    primaryClass = "astro-ph.CO",
    doi = "10.1007/s11433-026-2991-0",
    journal = "Sci. China Phys. Mech. Astron.",
    volume = "69",
    pages = "280413",
    year = "2026"
}

@article{Martin:2014nya,
    author = "Martin, Jerome and Ringeval, Christophe and Vennin, Vincent",
    title = "{Observing Inflationary Reheating}",
    eprint = "1410.7958",
    archivePrefix = "arXiv",
    primaryClass = "astro-ph.CO",
    doi = "10.1103/PhysRevLett.114.081303",
    journal = "Phys. Rev. Lett.",
    volume = "114",
    number = "8",
    pages = "081303",
    year = "2015"
}

@article{Mohammadi:2015jka,
    author = "Mohammadi, A. and Ossoulian, Z. and Golanbari, T. and Saaidi, K.",
    title = "{Intermediate inflation with modified kinetic term}",
    doi = "10.1007/s10509-015-2458-5",
    journal = "Astrophys. Space Sci.",
    volume = "359",
    number = "1",
    pages = "7",
    year = "2015"
}

@article{Mohammadi:2019qeu,
    author = "Mohammadi, Abolhassan and Golanbari, Tayeb and Saaidi, Khaled",
    title = "{Beta-function formalism for k-essence constant-roll inflation}",
    eprint = "1912.07006",
    archivePrefix = "arXiv",
    primaryClass = "gr-qc",
    doi = "10.1016/j.dark.2020.100505",
    journal = "Phys. Dark Univ.",
    volume = "28",
    pages = "100505",
    year = "2020"
}

@article{Mohammadi:2023kzd,
    author = "Mohammadi, Abolhassan and Kheirandish, Fardin",
    title = "{Exploring new subclass of k-inflation: Tachyon inflation in R+{\ensuremath{\eta}}T gravity model}",
    doi = "10.1016/j.dark.2023.101362",
    journal = "Phys. Dark Univ.",
    volume = "42",
    pages = "101362",
    year = "2023"
}

@article{Mohammadi:2025gbu,
    author = "Mohammadi, Abolhassan and Yogesh and Wang, Anzhong",
    title = "{Power law plateau inflation and primordial gravitational waves in the light of ACT}",
    eprint = "2507.06544",
    archivePrefix = "arXiv",
    primaryClass = "astro-ph.CO",
    doi = "10.1016/j.physletb.2025.140054",
    journal = "Phys. Lett. B",
    volume = "872",
    pages = "140054",
    year = "2026"
}

@article{Odintsov:2025eiv,
    author = "Odintsov, S. D. and Oikonomou, V. K.",
    title = "{Power-law F(R) gravity as deformations to Starobinsky inflation in view of ACT}",
    eprint = "2509.06251",
    archivePrefix = "arXiv",
    primaryClass = "gr-qc",
    doi = "10.1016/j.physletb.2025.139907",
    journal = "Phys. Lett. B",
    volume = "870",
    pages = "139907",
    year = "2025"
}

@unpublished{Odintsov:2025wai,
    author = "Odintsov, S. D. and Oikonomou, V. K.",
    title = "{GW170817 Viable Einstein-Gauss-Bonnet Inflation Compatible with the Atacama Cosmology Telescope Data}",
    eprint = "2506.08193",
    archivePrefix = "arXiv",
    primaryClass = "gr-qc",
    month = "6",
    year = "2025"
}

@article{Planck:2018jri,
    author = "Akrami, Y. and others",
    collaboration = "Planck",
    title = "{Planck 2018 results. X. Constraints on inflation}",
    eprint = "1807.06211",
    archivePrefix = "arXiv",
    primaryClass = "astro-ph.CO",
    doi = "10.1051/0004-6361/201833887",
    journal = "Astron. Astrophys.",
    volume = "641",
    pages = "A10",
    year = "2020"
}

@article{Qiu:2025iqm,
    author = "Qiu, Zhichong and Pang, Yehuang and Huang, Qingguo",
    title = "{The implications of inflation for the last ACT}",
    eprint = "2510.18320",
    archivePrefix = "arXiv",
    primaryClass = "astro-ph.CO",
    doi = "10.1007/s11433-025-2934-8",
    journal = "Sci. China Phys. Mech. Astron.",
    volume = "69",
    number = "6",
    pages = "260413",
    year = "2026"
}

@article{Remmen:2013eja,
    author = "Remmen, Grant N. and Carroll, Sean M.",
    title = "{Attractor Solutions in Scalar-Field Cosmology}",
    eprint = "1309.2611",
    archivePrefix = "arXiv",
    primaryClass = "gr-qc",
    reportNumber = "CALT-68-2853",
    doi = "10.1103/PhysRevD.88.083518",
    journal = "Phys. Rev. D",
    volume = "88",
    pages = "083518",
    year = "2013"
}

@article{SPT-3G:2025bzu,
    author = "Camphuis, E. and others",
    collaboration = "SPT-3G",
    title = "{SPT-3G D1: CMB temperature and polarization power spectra and cosmology from 2019 and 2020 observations of the SPT-3G main field}",
    eprint = "2506.20707",
    archivePrefix = "arXiv",
    primaryClass = "astro-ph.CO",
    reportNumber = "FERMILAB-PUB-25-0144-PPD",
    doi = "10.1103/7wt3-9v2y",
    journal = "Phys. Rev. D",
    volume = "113",
    number = "8",
    pages = "083504",
    year = "2026"
}

@unpublished{Sabogal:2026qvy,
    author = "Sabogal, Miguel A. and Iovino, Antonio J. and Vagnozzi, Sunny",
    title = "{Running into tension: primordial black holes from ultra-slow-roll inflation, spectral running, and the Hubble tension}",
    eprint = "2606.31362",
    archivePrefix = "arXiv",
    primaryClass = "astro-ph.CO",
    month = "6",
    year = "2026"
}

@article{Saikawa:2018rcs,
    author = "Saikawa, Ken'ichi and Shirai, Satoshi",
    title = "{Primordial gravitational waves, precisely: The role of thermodynamics in the Standard Model}",
    eprint = "1803.01038",
    archivePrefix = "arXiv",
    primaryClass = "hep-ph",
    reportNumber = "IPMU18-0037, MPP-2018-19",
    doi = "10.1088/1475-7516/2018/05/035",
    journal = "JCAP",
    volume = "05",
    pages = "035",
    year = "2018"
}

@article{Salopek:1992qy,
      author         = "Salopek, D. S. and Stewart, J. M.",
      title          = "{Hamilton-Jacobi theory for general relativity with
                        matter fields}",
      journal        = "Class. Quant. Grav.",
      volume         = "9",
      year           = "1992",
      pages          = "1943-1968",
      doi            = "10.1088/0264-9381/9/8/015",
      reportNumber   = "DAMTP-R-92-09",
      SLACcitation   = "%%CITATION = CQGRD,9,1943;%%"
}

@article{Salvio:2025izr,
    author = "Salvio, Alberto",
    title = "{Independent connection in action during inflation}",
    eprint = "2504.10488",
    archivePrefix = "arXiv",
    primaryClass = "hep-ph",
    doi = "10.1103/tq3v-vy3y",
    journal = "Phys. Rev. D",
    volume = "112",
    number = "6",
    pages = "L061301",
    year = "2025"
}

@unpublished{Shobcha:2026mpc,
    author = "Shobcha, Nehla and Sidik Risdianto, Norma and Budhi, Romy Hanang Setya",
    title = "{Minimal Extensions of the $\alpha$-Starobinsky Model: Reconciling ACT DR6 and Reheating Constraints}",
    eprint = "2606.24131",
    archivePrefix = "arXiv",
    primaryClass = "astro-ph.CO",
    month = "6",
    year = "2026"
}

@article{Solbi:2021rse,
    author = "Solbi, Milad and Karami, Kayoomars",
    title = "{Primordial black holes formation in the inflationary model with field-dependent kinetic term for quartic and natural potentials}",
    eprint = "2106.02863",
    archivePrefix = "arXiv",
    primaryClass = "astro-ph.CO",
    doi = "10.1140/epjc/s10052-021-09690-9",
    journal = "Eur. Phys. J. C",
    volume = "81",
    number = "10",
    pages = "884",
    year = "2021"
}

@article{Starobinsky:1982ee,
    author = "Starobinsky, Alexei A.",
    title = "{Dynamics of Phase Transition in the New Inflationary Universe Scenario and Generation of Perturbations}",
    doi = "10.1016/0370-2693(82)90541-X",
    journal = "Phys. Lett. B",
    volume = "117",
    pages = "175--178",
    year = "1982"
}

@article{Turner:1983he,
    author = "Turner, Michael S.",
    title = "{Coherent Scalar Field Oscillations in an Expanding Universe}",
    reportNumber = "EFI-83-29-CHICAGO",
    doi = "10.1103/PhysRevD.28.1243",
    journal = "Phys. Rev. D",
    volume = "28",
    pages = "1243",
    year = "1983"
}

@article{WMAP:2010qai,
    author = "Komatsu, E. and others",
    collaboration = "WMAP",
    title = "{Seven-Year Wilkinson Microwave Anisotropy Probe (WMAP) Observations: Cosmological Interpretation}",
    eprint = "1001.4538",
    archivePrefix = "arXiv",
    primaryClass = "astro-ph.CO",
    doi = "10.1088/0067-0049/192/2/18",
    journal = "Astrophys. J. Suppl.",
    volume = "192",
    pages = "18",
    year = "2011"
}

@article{Watanabe:2006qe,
    author = "Watanabe, Yuki and Komatsu, Eiichiro",
    title = "{Improved Calculation of the Primordial Gravitational Wave Spectrum in the Standard Model}",
    eprint = "astro-ph/0604176",
    archivePrefix = "arXiv",
    doi = "10.1103/PhysRevD.73.123515",
    journal = "Phys. Rev. D",
    volume = "73",
    pages = "123515",
    year = "2006"
}

@article{Yin:2025rrs,
    author = "Yin, Wen",
    title = "{Higgs-like inflation ACTivated mass}",
    eprint = "2505.03004",
    archivePrefix = "arXiv",
    primaryClass = "hep-ph",
    doi = "10.1088/1475-7516/2025/09/062",
    journal = "JCAP",
    volume = "09",
    pages = "062",
    year = "2025",
    number         = {},
}

@unpublished{Yogesh:2025wak,
    author = "Yogesh and Mohammadi, Abolhassan and Wu, Qiang and Zhu, Tao",
    title = "{Starobinsky like inflation and EGB Gravity in the light of ACT}",
    eprint = "2505.05363",
    archivePrefix = "arXiv",
    primaryClass = "astro-ph.CO",
    month = "5",
    year = "2025"
}

@article{Zahoor:2025nuq,
    author = "Zahoor, Mehnaz and Khan, Suhail and Bhat, Imtiyaz Ahmad",
    title = "{Reconciling fractional power potential and EGB gravity in the light of ACT}",
    eprint = "2507.18684",
    archivePrefix = "arXiv",
    primaryClass = "astro-ph.CO",
    doi = "10.1016/j.jheap.2025.100458",
    journal = "JHEAp",
    volume = "49",
    pages = "100458",
    year = "2026"
}

@article{Zharov:2025zjg,
    author = "Zharov, D. S. and Sobol, O. O. and Vilchinskii, S. I.",
    title = "{ACT observations, reheating, and Starobinsky and Higgs inflation}",
    eprint = "2505.01129",
    archivePrefix = "arXiv",
    primaryClass = "astro-ph.CO",
    doi = "10.1103/km3q-rm34",
    journal = "Phys. Rev. D",
    volume = "112",
    number = "2",
    pages = "023544",
    year = "2025"
}

@article{starobinsky:1980te,
    author = "Starobinsky, Alexei A.",
    editor = "Khalatnikov, I. M. and Mineev, V. P.",
    title = "{A New Type of Isotropic Cosmological Models Without Singularity}",
    doi = "10.1016/0370-2693(80)90670-X",
    journal = "Phys. Lett. B",
    volume = "91",
    pages = "99--102",
    year = "1980"
}

@article{Albrecht:1982mp,
    author = "Albrecht, Andreas and Steinhardt, Paul J. and Turner, Michael S. and Wilczek, Frank",
    title = "{Reheating an Inflationary Universe}",
    reportNumber = "UPR-0189T, EFI-82-09-CHICAGO",
    doi = "10.1103/PhysRevLett.48.1437",
    journal = "Phys. Rev. Lett.",
    volume = "48",
    pages = "1437",
    year = "1982"
}

@article{Kofman:1994rk,
    author = "Kofman, Lev and Linde, Andrei D. and Starobinsky, Alexei A.",
    title = "{Reheating after inflation}",
    eprint = "hep-th/9405187",
    archivePrefix = "arXiv",
    reportNumber = "UH-IFA-94-35, SU-ITP-94-13, YITP-U-94-15",
    doi = "10.1103/PhysRevLett.73.3195",
    journal = "Phys. Rev. Lett.",
    volume = "73",
    pages = "3195--3198",
    year = "1994"
}

@article{Kofman:1997yn,
    author = "Kofman, Lev and Linde, Andrei D. and Starobinsky, Alexei A.",
    title = "{Towards the theory of reheating after inflation}",
    eprint = "hep-ph/9704452",
    archivePrefix = "arXiv",
    reportNumber = "IFA-97-28, SU-ITP-97-18",
    doi = "10.1103/PhysRevD.56.3258",
    journal = "Phys. Rev. D",
    volume = "56",
    pages = "3258--3295",
    year = "1997"
}

@article{Bassett:2005xm,
    author = "Bassett, Bruce A. and Tsujikawa, Shinji and Wands, David",
    title = "{Inflation dynamics and reheating}",
    eprint = "astro-ph/0507632",
    archivePrefix = "arXiv",
    doi = "10.1103/RevModPhys.78.537",
    journal = "Rev. Mod. Phys.",
    volume = "78",
    pages = "537--589",
    year = "2006"
}

@article{Rehagen:2015zma,
    author = "Rehagen, Thomas and Gelmini, Graciela B.",
    title = "{Low reheating temperatures in monomial and binomial inflationary potentials}",
    eprint = "1504.03768",
    archivePrefix = "arXiv",
    primaryClass = "hep-ph",
    doi = "10.1088/1475-7516/2015/06/039",
    journal = "JCAP",
    volume = "06",
    pages = "039",
    year = "2015",
    number         = {},
}

@unpublished{Bhat:2020wdc,
    author = "Bhat, Imtiyaz Ahmad and Chakravarty, Girish Kumar and Adhikari, Rathin",
    title = "{Inflation, reheating, leptogenesis and bounds on soft supersymmetry breaking parameters}",
    eprint = "2012.15256",
    archivePrefix = "arXiv",
    primaryClass = "hep-ph",
    month = "12",
    year = "2020"
}

@article{Bezerra-Sobrinho:2025gfg,
    author = "Bezerra-Sobrinho, J. and Medeiros, L. G.",
    title = "{Starobinsky inflation and the latest CMB data: a subtle tension?}",
    eprint = "2511.06640",
    archivePrefix = "arXiv",
    primaryClass = "astro-ph.CO",
    doi = "10.1140/epjc/s10052-026-15686-0",
    journal = "Eur. Phys. J. C",
    volume = "86",
    number = "4",
    pages = "416",
    year = "2026"
}

@article{Babichev:2007dw,
    author = "Babichev, Eugeny and Mukhanov, Viatcheslav and Vikman, Alexander",
    title = "{k-Essence, superluminal propagation, causality and emergent geometry}",
    eprint = "0708.0561",
    archivePrefix = "arXiv",
    primaryClass = "hep-th",
    reportNumber = "LMU-ASC-54-07",
    doi = "10.1088/1126-6708/2008/02/101",
    journal = "JHEP",
    volume = "02",
    pages = "101",
    year = "2008"
}

@article{Deffayet:2010qz,
    author = "Deffayet, Cedric and Pujolas, Oriol and Sawicki, Ignacy and Vikman, Alexander",
    title = "{Imperfect Dark Energy from Kinetic Gravity Braiding}",
    eprint = "1008.0048",
    archivePrefix = "arXiv",
    primaryClass = "hep-th",
    reportNumber = "CERN-PH-TH-2010-166",
    doi = "10.1088/1475-7516/2010/10/026",
    journal = "JCAP",
    volume = "10",
    pages = "026",
    year = "2010"
}

@article{Odintsov:2026dss,
    author = "Odintsov, S. D. and Oikonomou, V. K.",
    title = "{ACT data and positive running of the spectral index for scalar theory and modified gravity}",
    eprint = "2605.17813",
    archivePrefix = "arXiv",
    primaryClass = "gr-qc",
    doi = "10.1016/j.physletb.2026.140805",
    journal = "Phys. Lett. B",
    volume = "880",
    pages = "140805",
    year = "2026"
}

@article{Oikonomou:2026mvp,
    author = "Oikonomou, V. K.",
    title = "{Analytic singular slow-roll inflation}",
    eprint = "2603.11794",
    archivePrefix = "arXiv",
    primaryClass = "gr-qc",
    doi = "10.1088/1361-6382/ae97d0",
    journal = "Class. Quant. Grav.",
    volume = "43",
    number = "16",
    pages = "165013",
    year = "2026"
}

@article{Odintsov:2026doe,
    author = "Odintsov, S. D. and Oikonomou, V. K.",
    title = "{R2-corrected Tachyon Scalar Field Inflation, the ACT Data, and Phantom Transition}",
    eprint = "2601.21364",
    archivePrefix = "arXiv",
    primaryClass = "gr-qc",
    doi = "10.1016/j.nuclphysb.2026.117384",
    journal = "Nucl. Phys. B",
    volume = "1025",
    pages = "117384",
    year = "2026"
}

@article{Oikonomou:2026vrw,
    author = "Oikonomou, V. K.",
    title = "{General scalar field inflation ACT attractors: Utilizing the ns(r) relation}",
    eprint = "2601.20053",
    archivePrefix = "arXiv",
    primaryClass = "gr-qc",
    doi = "10.1016/j.physletb.2026.140430",
    journal = "Phys. Lett. B",
    volume = "876",
    pages = "140430",
    year = "2026"
}

@article{Oikonomou:2025htz,
    author = "Oikonomou, V. K.",
    title = "{Strong gravity effects on R2-corrected single scalar field inflation and compatibility with the ACT data}",
    eprint = "2508.17363",
    archivePrefix = "arXiv",
    primaryClass = "gr-qc",
    doi = "10.1016/j.physletb.2025.139972",
    journal = "Phys. Lett. B",
    volume = "871",
    pages = "139972",
    year = "2025"
}

\end{document}